\documentclass[%
 reprint,
superscriptaddress,
 amsmath,amssymb,
 aps,
pra,
]{revtex4-2}

\usepackage{graphicx}
\usepackage{dcolumn}
\usepackage{bm}
\usepackage{braket}
\usepackage{siunitx}
\usepackage{hyperref}
\usepackage{xr}

\begin{document}

\makeatletter
\def\@fnsymbol#1{\ensuremath{\ifcase#1\or *\or \dagger\or \ddagger\or
   \mathsection\or \mathparagraph\or \|\or **\or \dagger\dagger
   \or \ddagger\ddagger \else\@ctrerr\fi}}
\def\@makefnmark{\hbox{\@textsuperscript{\normalfont\@thefnmark}}}
\def\@makefntext#1{\parindent 1em\noindent
   \hb@xt@1.8em{%
     \hss\@textsuperscript{\normalfont\@thefnmark}}#1}
\makeatother

\vspace{-0.2cm}

\author{Molly A. Thomas}
\email{molly.a.thomas@ucl.ac.uk}
\affiliation{Quantum Engineering Technology Labs, H. H. Wills Physics Laboratory and School of Electrical, Electronic, and Mechanical Engineering, University of Bristol, BS8 1FD, UK}
\affiliation{Department of Electronic and Electrical Engineering, University College London, WC1E 7JE, UK}
\thanks{Current affiliation}

\author{Daniel Llewellyn}
\affiliation{Quantum Engineering Technology Labs, H. H. Wills Physics Laboratory and School of Electrical, Electronic, and Mechanical Engineering, University of Bristol, BS8 1FD, UK}

\author{Patrick W. Yard}
\affiliation{Quantum Engineering Technology Labs, H. H. Wills Physics Laboratory and School of Electrical, Electronic, and Mechanical Engineering, University of Bristol, BS8 1FD, UK}

\author{Benjamin A. Slater}
\affiliation{Quantum Engineering Technology Labs, H. H. Wills Physics Laboratory and School of Electrical, Electronic, and Mechanical Engineering, University of Bristol, BS8 1FD, UK}

\author{Caterina Vigliar}
\affiliation{Quantum Engineering Technology Labs, H. H. Wills Physics Laboratory and School of Electrical, Electronic, and Mechanical Engineering, University of Bristol, BS8 1FD, UK}
\affiliation{Department of Electrical and Photonics Engineering, Technical University of Denmark, 2800 Kgs. Lyngby, Denmark}

\author{Stefano Paesani}
\affiliation{Quantum Engineering Technology Labs, H. H. Wills Physics Laboratory and School of Electrical, Electronic, and Mechanical Engineering, University of Bristol, BS8 1FD, UK}
\affiliation{NNF Quantum Computing Programme, Niels Bohr Institute, University of Copenhagen, Blegdamsvej 17, 2100 Copenhagen, Denmark}
\thanks{Current affiliation}

\author{Massimo Borghi}
\affiliation{Quantum Engineering Technology Labs, H. H. Wills Physics Laboratory and School of Electrical, Electronic, and Mechanical Engineering, University of Bristol, BS8 1FD, UK}
\affiliation{Dipartimento di Fisica, Università di Pavia, Via Agostino Bassi 6, 27100 Pavia, Italy}
\thanks{Current affiliation}

\author{D{\"o}nd{\"u} {\c{S}}ahin}
\affiliation{Quantum Engineering Technology Labs, H. H. Wills Physics Laboratory and School of Electrical, Electronic, and Mechanical Engineering, University of Bristol, BS8 1FD, UK}
\affiliation{The Boeing Company, Tukwila, Washington 98108, United States of America}
\thanks{Current affiliation}

\author{John G. Rarity}
\affiliation{Quantum Engineering Technology Labs, H. H. Wills Physics Laboratory and School of Electrical, Electronic, and Mechanical Engineering, University of Bristol, BS8 1FD, UK}

\author{Leif K. Oxenl\o we}
\affiliation{Department of Electrical and Photonics Engineering, Technical University of Denmark, 2800 Kgs. Lyngby, Denmark}

\author{Mark G. Thompson}
\affiliation{Quantum Engineering Technology Labs, H. H. Wills Physics Laboratory and School of Electrical, Electronic, and Mechanical Engineering, University of Bristol, BS8 1FD, UK}

\author{Karsten Rottwitt}
\affiliation{Department of Electrical and Photonics Engineering, Technical University of Denmark, 2800 Kgs. Lyngby, Denmark}

\author{Yunhong Ding}
\affiliation{Department of Electrical and Photonics Engineering, Technical University of Denmark, 2800 Kgs. Lyngby, Denmark}

\author{Jianwei Wang}
\affiliation{Quantum Engineering Technology Labs, H. H. Wills Physics Laboratory and School of Electrical, Electronic, and Mechanical Engineering, University of Bristol, BS8 1FD, UK}
\affiliation{State Key Laboratory for Mesoscopic Physics, School of Physics, Peking University, Beijing, 100871, China}
\thanks{Current affiliation}

\author{Davide Bacco}
\affiliation{Department of Electrical and Photonics Engineering, Technical University of Denmark, 2800 Kgs. Lyngby, Denmark}
\affiliation{Department of Physics and Astronomy, University of Florence, 50019, Firenze, Italy}
\thanks{Current affiliation}

\author{Jorge Barreto}
\email{ghk.barreto@gmail.com}
\affiliation{Quantum Engineering Technology Labs, H. H. Wills Physics Laboratory and School of Electrical, Electronic, and Mechanical Engineering, University of Bristol, BS8 1FD, UK}

\vspace{-0.2cm}

\title{\large High-dimensional Path-Encoded Entanglement Distribution Between Photonic Chips Enabled by Multimode Phase Stabilisation} 

\vspace{0.3cm}

\begin{abstract}
    The reliable distribution of high-dimensional entangled quantum states, an important resource in quantum technologies, through optical fibre networks is challenging due to the need to maintain coherence across multiple modes. 
    Here we demonstrate the distribution of four-dimensional path-encoded entangled quantum states between photonic chips, enabled by a novel multimode phase stabilisation algorithm. The algorithm utilises the reconfigurability of the integrated photonic circuits to complete one iteration of phase stabilisation in just two measurement rounds for an arbitrary number of modes, and requires no additional hardware to the quantum measurements it enables.
    As a result, we are able to perform complete quantum state tomography across two chips using the minimum number of local projective measurements to verify the fidelity of the distributed entangled state to be $86\%$ (compared to $8.1\%$ without the phase stabilisation) with an entanglement entropy of $0.995\pm0.002$.
\end{abstract}

\maketitle


\section*{Introduction}

Entanglement is a cornerstone of quantum technologies with no classical counterpart, and its distribution is essential for applications including entanglement-based quantum key distribution \cite{Yin2020-EntQKDBenefits, Koashi2003-SourceIndepSecu}, networked quantum computing \cite{Psi2024-Hardware, PhotonInc2024-DisQComp, Knorzer2025-DistributedQIP}, and distributed quantum sensing \cite{Kim2024-DisQSensing,Guo2020-DisQSensingCV}. Key limiting factors in widely deploying distributed entanglement are achievable communication distances \cite{Neumann2022-EntDisDistance, Leent2022-EntDisDistance}, and the reliable distribution of quantum states at high rates \cite{Ecker2021-EntDisRates,Mueller2024-EntDisRates,Zhuang2025-EntDisRates}.
One way to increase the distribution rate of quantum information is by employing high-dimensional (HD) quantum states \cite{Erhard2020-HDEntReview}, qudits, where information is encoded in more than two modes per particle. Such states have greater information capacity per carrier 
\cite{Cerf2002-QKDSecurityDLevel}, greater noise resilience \cite{Ecker2019-HDNoiseRobust, Cozzolino2019-HDQCommBenefits}, and less overheads for error correction \cite{Vigliar2021-QubitError, Chi2022-QuditQComp} than their binary counterparts, qubits.

\begin{figure*}
  \includegraphics[width=1\textwidth]{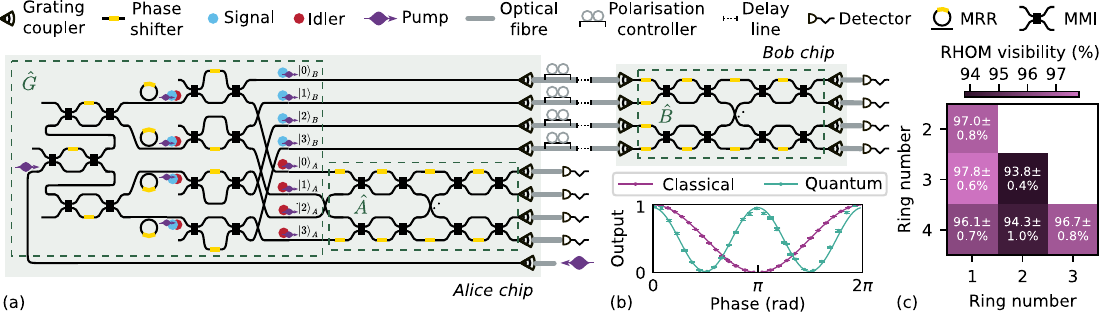}
  \caption{(a) Schematic of the experimental set-up consisting of two integrated photonic circuits, Alice and Bob, connected by optical fibres. The pairs of entangled qudits are prepared in section $\hat{G}$ of the Alice chip and projective measurements are performed by Alice and Bob with the reconfigurable interferometers $\hat{A}$ and $\hat{B}$. (b) Interference fringes in optical power (classical) and coincidence counts (quantum) due to time reversed Hong-Ou-Mandel interference. (c) RHOM interference visibilities of all ring pairs, quantifying indistinguishability.
}
  \label{fig:setup}
\end{figure*}

Photons represent a natural carrier for distributing quantum states due to their limited decoherence compared to matter systems, room temperature operation, and the availability of mature, low-loss fibre optic technology \cite{Romero2024-PhotonicQC}.
Integrated photonics can be used to encode qudits in a variety of degrees of freedom (DoF) such as orbital angular momentum (OAM) \cite{Mair2001-OAMEntEncoding, Cozzolino2019-OAMHD}, time \cite{Cheng2023-HDEntTimeFreq, Bensemhoun2024-HDEntTimeFreq}, frequency \cite{Kues2017-HDEntFreq}, and path \cite{Wang2018-16D, DaLio2021-PathQKD2kmMCF, Bacco2017-SpaceMultiplexingChipToChip}. There are trade-offs with all DoF when considering ease of generation, scalability, manipulation, and stability \cite{Wang2020-IntegratedQPhotonics}.
As OAM and spatial modes are spatio-temporally degenerate, fluctuations in the environment can lead to coupling between modes making long distance distribution difficult \cite{Ni2025-OAMCrosstalk}. Systems using continuous degrees of freedom such as time or frequency can be designed with larger separation between modes to minimise cross-talk. However, implementing gates on these systems requires high speed electronic and photonic control and delay lines, or frequency converters which can both be experimentally challenging to implement.
An alternative to encoding HD information using more modes in one DoF is to use a smaller number of modes in multiple DoF: hybrid encoding \cite{Steinlechner2017-HyperEnt}. For example, ququart (four-dimensional qudit) hybrid encoding has been demonstrated by combining path and polarisation modes \cite{Ciampini2016-HyperEnt}; polarisation and transverse modes \cite{Zheng2023-HDEntDisChips}; time and frequency modes \cite{Congia2025-OnChipHyperentangled}; and path and transverse modes \cite{Forbes2025-HybridEnc}.
However, accessible state space and fidelities can be limited with this method as they inherit the experimental challenges of implementing each DoF.
On the other hand, path encoding - where information is encoded in physically distinct paths representing different modes - is a natural way of encoding information as linear optic operations are performed by phase shifters and beamsplitters, which act on one and
two path modes \cite{Knill2001-LinearOpticQC}.
As such, path encoding is often essential for manipulation and measurement of photonic states, meaning conversion between other degrees of freedom to path is required at the expense of loss and errors \cite{Llewellyn2019-Tele,Zheng2023-HDEntDisChips}.

While integrated photonics is inherently phase stable, the fibre links over which quantum states are distributed induce phase noise in states due to thermal and mechanical fluctuations affecting each mode differently \cite{DaLio2021-PathQKD2kmMCF,DaLio2020-2kmStableTransmission}.
This makes the distribution of HD quantum states challenging, regardless of the encoding chosen, due to the need to maintain coherence across multiple modes through a noisy distribution link. The distribution of path-encoded entangled qudits between chips has thus far been precluded due to the difficulty in maintaining phase coherence of more than two path-modes at once \cite{Zheng2023-HDEntDisChips, Bacco2017-SpaceMultiplexingChipToChip}. Existing schemes for stabilisation using phase-locked loops, typically have a large overhead, requiring additional FGPAs and phase modulators on each mode \cite{Larsen2025-XanaduOnChipGKP, DaLio2021-PathQKD2kmMCF}. They are also difficult to implement when stabilising more than two modes as the reference signal is dependent on multiple parameters.

Here we demonstrate the distribution of HD path-encoded entangled states between two photonic chips, enabled by a multimode active phase stabilisation algorithm. This method is valid for the stabilisation of an arbitrary number of modes and does not require additional hardware to the quantum measurements it enables.
We validate our approach by performing quantum state tomography across the two chips using local measurements on the entangled photons, yielding a fidelity of the entangled ququart state of 86\%. We do this using the minimum set of single-qudit measurements required for complete tomography of a HD state, only possible due to the coherence of all modes in the transmission channel, as opposed to just pairs of modes.

\section*{Experimental set-up}

Our system generates and measures quantum states distributed between two silicon-on-insulator photonic chips – Alice and Bob (Fig. \ref{fig:setup}a). 
Alice contains micro-ring resonator (MRR) sources which we pump with a continuous-wave laser to probabilistically produce non-degenerate entangled photon pairs via spontaneous four-wave mixing (SFWM) \cite{Baboux2023-AlGaAsQPhotonics, Borghi2017SiNonLinear}.
At the output of the sources, each co-propagating signal and idler photon pair is spatially separated with an asymmetric-Mach-Zehnder interferometer (AMZI) acting as an on-chip filter. As a result we have four signal path modes and four idler path modes at the output of the $\hat{G}$ section of the transmitter chip, as can be seen in Fig. \ref{fig:setup}a.

The signal photons are coupled off-chip via grating couplers to a fibre array, and distributed to the Bob chip. On each path mode of this inter-chip link we add a polarisation controller, to maximise optical coupling to Bob, and a tunable delay line, to path-length match each mode in order to maintain temporal coherence.
Photon loss in the inter-chip link affects the system in several ways. Uniform loss across all modes only affects the distribution rate of the state, however, if the loss is non-uniform across the modes the relative amplitudes in the superposition are altered, reducing the state fidelity.
To minimise this effect, on-chip attenuators on each mode balance the relative losses of the paths at the inputs of Bob.

At this stage, we probe the quantum state of the idler and signal photons using Alice and Bob's reconfigurable Mach-Zehnder interferometer (MZI) networks to implement the operators $\hat{A}$ and $\hat{B}$. This allows us to perform projective measurements on the entangled quantum state and extract the fidelity of its distribution.
Finally, the signal and idler photons are coupled off-chip to be measured by superconducting-nanowire single photon detectors (SNSPDs) in order to collate statistics on the projective measurement outcomes. More details on the characterisation of the experimental set-up can be found in the Supplementary Information.

In order to validate our approach, we choose to prepare maximally-entangled two-qudit states. Simultaneously and coherently pumping $d$ MRRs in the low photon-number regime (such that we assume at most one photon pair is generated) results in an entangled two-photon state of dimension $d$ \cite{Wang2018-16D}, which can be described as
\begin{equation}
    \ket{\Phi_d} = \sum_{n=0}^{d-1}\lvert \alpha_{n}\rvert e^{i\phi_{n}}\ket{nn},
\end{equation}
where $n$ denotes the path mode, $\lvert \alpha_{n}\rvert$ is the probability of a photon pair being in the $n$-th path mode, and $\phi_{n}$ is the phase of the $n$-th path mode.
The $\phi_{n}$ phases are controlled by tuneable thermo-optic phase shifters (TOPS) and the $\lvert \alpha_{n}\rvert$ amplitudes are controlled by changing the optical power with which the sources are pumped using tunable beamsplitters, in the form of MZIs with TOPS, before the sources. A HD Bell-like state is produced when $\alpha_n=\frac{1}{\sqrt{d}}$. Pumping our four MRRs, therefore, produces two entangled, path-encoded $d=4$ qudits (ququart).
We use the visibility of time-reversed Hong-Ou-Mandel (RHOM) interference fringes to quantify the indistinguishability of pairs of our sources \cite{Chen2007-RHOM,Silverstone2014-OnChipInterference}. A fringe for one pair can be seen in Fig. \ref{fig:setup}b, with visibilities of RHOM interference fringes between all pairs of rings shown in Fig. \ref{fig:setup}b. 
Further details of the source and RHOM measurements are reported in the Supplementary Information.

\begin{figure*}
\centering
\includegraphics[width=1\textwidth]{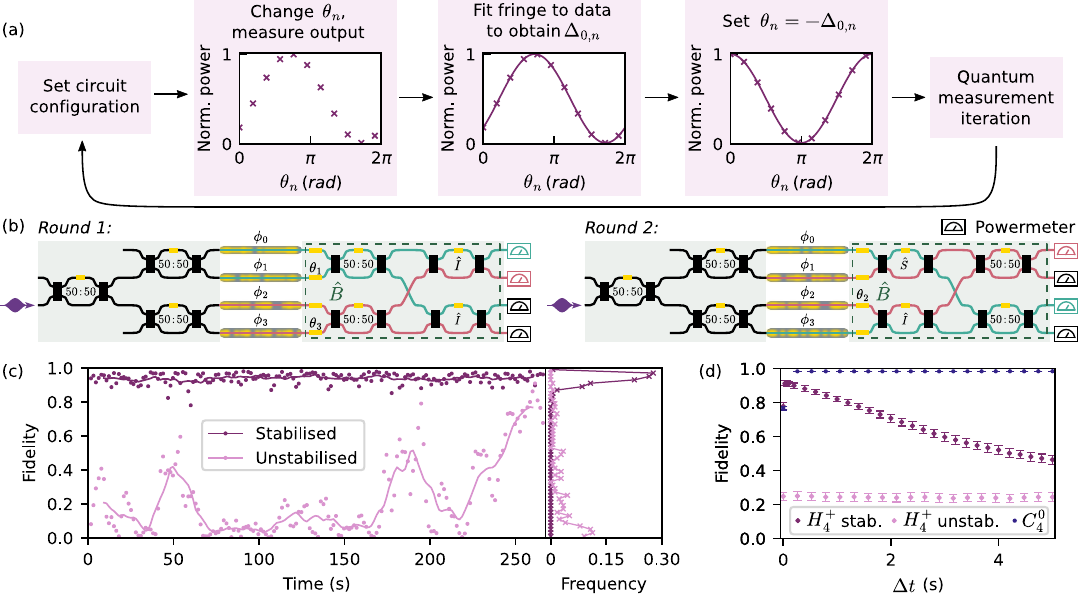}
\caption{(a) Workflow of measurement of the relative phase between two modes, interleaved with iterations of the desired quantum measurement. (b) Interferometer configurations for two rounds of the phase stabilisation of our four-dimensional system, where MZIs are set to 50:50; identity, $\hat{I}$; or swap, $\hat{S}$, to interfere different pairs of modes and route light to off-chip powermeters. The powermeters highlighted in green and pink are those used to measure fringes in the phase stabilisation. In round one we measure $\Delta_{0,1}$ and $\Delta_{2,3}$ and in round two we measure $\Delta_{0,3}$ and $\Delta_{1,2}$. (c) Fidelity of the classical $H_4^+$ Hadamard basis state measurement over time without and with phase stabilisation. (d) Fidelity decay at time $\Delta t$ since stabilisation - an iteration of the phase stabilisation is complete at $\Delta t =0s$. This is compared to the fidelity of the computational basis state $C_4^0=(1 \,\, 0 \,\, 0\,\, 0)^T$.
}
\label{fig:phase_stab}
\end{figure*}

\section*{Phase stabilisation in the inter-chip channels}

In order to maintain the correct superposition of modes in a quantum state, relative phases between the modes must remain coherent. While relative phase remains stable between on-chip path modes, this is not the case when path modes propagate through optical fibres \cite{DaLio2021-PathQKD2kmMCF,DaLio2020-2kmStableTransmission}. 
As such, active phase compensation is essential to maintain phase coherence as the photons are transmitted through our fibre link.

\begin{figure*}
\includegraphics[width=\textwidth]{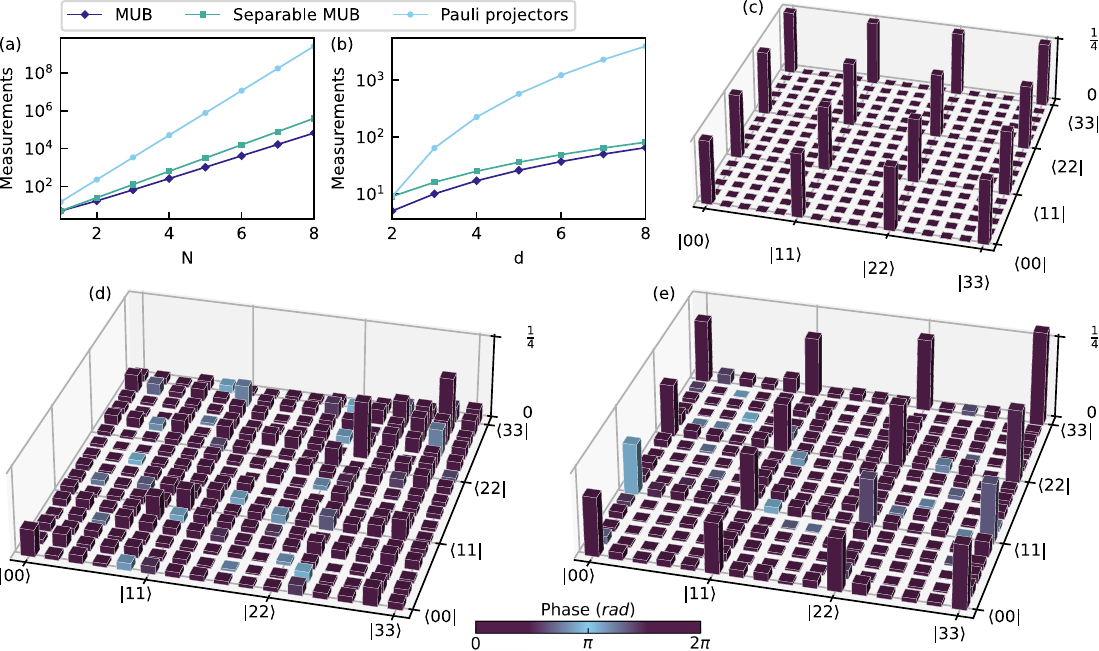}
\caption{(a) Number of projective measurements required for $N$ qudits with dimension $d=4$ (b) Number of projective measurements required for $N=2$ qudits with dimension $d$. (c) Ideal, (d) non-stabilised, and (e) stabilised reconstructed density matrix for the experimentally prepared, distributed, and measured $\ket{\Phi^+_4}$ state.
}
\label{fig:quantum_results}
\end{figure*}

As only relative phases need to be corrected, we select one path, labelled mode 0, as a phase reference, against which all other phases are corrected.
We can measure the phase of a mode $n$ relative to the reference, $\Delta_{0,n}=\phi_n-\phi_0$, by constructing a MZI between the modes across the two chips by reconfiguring Bob's interferometer. We actively drive phase changes on mode $n$, $\theta_n$, by applying an electrical power, $E_n$, to the TOPS on that mode to observe the phase-dependent fringe in optical output
\begin{equation}
    P(\theta_n) = P_{max}\cos^2(\nu(E_n-E_{0,n})) + P_{min},
    \label{eq:fringe}
\end{equation}
where $P_{max}$ and $P_{min}$ are the maximum and minimum optical powers in the fringe, $E_{0,n}$ is the electrical power corresponding to $\Delta_{0,n}$, and $\nu$ is the frequency relating electrical power to phase. We fit Eq. \ref{eq:fringe} to the measured optical power fringe to obtain its phase offset, ${\Delta_{0,n}=-\nu E_{0,n}}$, which we can then compensate for by setting the phase shifter on that mode to $\theta_{n} = -\Delta_{0,n}$ at the input of the receiver. The process for measurement of phase drift between two modes can be seen in Fig. \ref{fig:phase_stab}a.

This approach can be extended to an arbitrarily large number of path modes, $d$, by sequentially interfering each of the $d-1$ modes with the reference to correct for $\Delta_{0,n}$. However, this would require $d-1$ rounds of measurement per iteration of phase stabilisation to stabilise the link. Alternatively, by simultaneously measuring pairs of modes and inferring phase offsets, we can stabilise any size system in just two rounds of measurement per iteration.
In round one we configure $\hat{B}$ to interfere pairs of modes $n$ and $n+1$ (where $n$ is even and mod $d-1$) to measure $\Delta_{n,n+1}=\phi_{n+1}-\phi_n$ by changing $\theta_{n+1}$. In a second round we configure $\hat{B}$ to interfere pairs modes $n+1$ and $n+2$ to measure $\Delta_{n+1,n+2}=\phi_{n+2}-\phi_{n+1}$ by changing $\theta_{n+2}$. We can then infer $\Delta_{0,n}=\Delta_{0,n-1}+\Delta_{n-1,n}$ (or $\Delta_{0,n}=\Delta_{0,n+1}+\Delta_{n+1,n}$) for modes not measured directly against the reference by combining measurements from both rounds to infer all $\Delta_{0,n}$ in just two rounds of measurement. Fig. \ref{fig:phase_stab}b shows the interferometer configurations for our $d=4$ system.

We achieve this phase stabilisation classically, using residue pump light sufficiently bright to be measured by photodiodes which, in our implementation, have a noise floor of \SI{-70}{dBm} to \SI{-60}{dBm}.
This pump light co-propagates with the signal photons in the noisy inter-chip link, therefore picking up the same phase drifts. We separate this pump light from the signal photons at the output of the receiver with off-chip band-pass filters in order to suppress pump leakage to the SNSPDs. We direct this pump wavelength light to powermeters for system characterisation classically and, therefore, also use it to implement our phase stabilisation algorithm. As such, this method uses no additional hardware to what is required for the quantum measurement it enables.

Fig. \ref{fig:phase_stab}c shows the fidelity of classical light in the state $H_4^+ = \tfrac{1}{2}(1\,\, 1\,\, 1\,\, 1)^T$ that has passed through the four-dimensional inter-chip interferometer without and with phase stabilisation. This demonstrates improvement in average fidelity from 21\% to 95\% and reduction in variance from 86\% to 3\% in this sample.

Iterations of the quantum measurement must be interleaved with the phase stabilisation, due to the need to reconfigure the chip to perform the phase stabilisation. This means that iterations of the phase stabilisation must be sufficiently fast to maximise quantum measurement acquisition time and ensure that the accuracy of the phase correction is maintained. This is illustrated in Fig. \ref{fig:phase_stab}(d) where the fidelity of the classical $H_4^+$ state degrades in the time, $\Delta t$, following completion of one iteration of phase stabilisation. 
An iteration of the phase stabilisation algorithm as-implemented takes \SI{0.94}{\s}, with 82\% of the time required being to set the on-chip phase shifters.
Such significant time is required due to the 0.03s clock cycle of the electronics which are used for multi-channel control of the integrated photonics.
This characterisation informs our choice of how to handle the trade-off between fidelity and acquisition time of our quantum measurement. As we prioritise the fidelity of quantum state distribution here, we choose to acquire the quantum measurement in iterations of \SI{0.2}{\s}, resulting in a duty cycle of 18$\%$.
Further details on the phase stabilisation algorithm and its implementation are reported in the Supplementary Information.

\section*{Quantum state distribution results}

We prepare the HD Bell-like state
\begin{equation}
    \ket{\Phi^+_4} = \frac{1}{2}(\ket{00}+\ket{11}+\ket{22}+\ket{33})
\end{equation}
in order to verify the reliable distribution of our path-encoded entangled ququarts. We perform quantum state tomography (QST) to estimate the density matrix that represents the experimentally prepared, distributed, and measured state. QST entails preparing many copies of the same state and performing measurements on these copies in different measurement bases, allowing us to construct a density matrix, $\rho$, that represents the experimental state \cite{Altepeter2005-QST}.

There are many different measurement basis choices that can be made to perform QST, each requiring a particular number of measurements to be an informationally complete basis.
When measuring a single qubit, the common choice is to measure the projectors of the Pauli operators \cite{Nielsen_Chuang}. When we expand to larger-scale systems with $N$ qubits and/or higher dimensions, the most efficient complete set of measurements are mutually unbiased (MUB) measurements, requiring $d^N+1$ measurements to span the full state space \cite{Wootters1989-MUBTomo}.
MUB measurements across a full multi-particle system require non-local measurements, however we must perform local measurements as our state consists of two photons that are physically disparate. A common method of locally measuring multi-photon states is to perform a complete set of combinations $d$-dimensional generalised Pauli (or Gell-Mann) operator \cite{Bertlmann2008-GellMannGen} projections on the individual photons \cite{Wang2018-16D,Bao2023-VeryLargeScale, Zheng2023-HDEntDisChips,Adcock2019-4PhotonGraphStates}. This requires $(d^2-1)^N$ projective measurements which quickly becomes prohibitive as system size increases. Alternatively to using generalised Pauli operators, we use local MUB measurements to probe the state of the individual photons. This requires $(d+1)^N$ projective measurements, constituting the minimum number of measurements required to perform complete tomography with local measurements \cite{Usenko2024-QSTMaths}. Local MUB measurements bring us far closer to the minimum bound imposed by joint MUB measurements as $N$ and $d$ increase, as shown in Fig. \ref{fig:quantum_results}a and b respectively. Further explanation of the construction of the quantum state tomography measurement and a practical guide is reported in the Supplementary Information.
These MUB measurements require all $d$ modes to be coherent, unlike $d$-dimensional Pauli operator projectors which only require coherence across pairs of modes (Supplementary Information). As such, we are only able to use the measurement-efficient MUB due to our implemented active phase stabilisation.

We quantify fidelity, $\mathcal{F} = \left(Tr\left(\sqrt{\sqrt{\rho}\tilde{\rho}\sqrt{\rho}} \right)\right)^2$, by comparing the measured density matrix, $\rho$, to the ideal density matrix, $\tilde{\rho}$ \cite{Altepeter2005-QST}. 
The ideal density matrix for the state $\ket{\Phi^+_4}$ can be seen in Fig. \ref{fig:quantum_results}c. The reconstructed density matrix from quantum state tomography of the experimentally prepared, distributed, and measured $\ket{\Phi^+_4}$ state without active phase stabilisation can be seen in Fig. \ref{fig:quantum_results}d with a fidelity of 8.1\%. Fig. \ref{fig:quantum_results}e shows the resulting density matrix when the active phase stabilisation is implemented, illustrating the significant increase in fidelity to 86\%. A high degree of entanglement is preserved, quantified by an entanglement entropy \cite{Altepeter2005-QST}, $\mathcal{E}=0.995\pm0.002$ (where $\mathcal{E}=1$  corresponds to a maximally entangled state). A dimension witness \cite{Wang2018-16D} verifies the dimension of the prepared ququarts to be 4. Both metrics are defined in the Supplementary Information.

\section*{Conclusion}
We have successfully demonstrated the generation and distribution of path entangled four-dimensional qudits between integrated photonic chips. This demonstrates the feasibility of distributing HD path-encoded quantum states between remote quantum processors. Key to this achievement is the novel scheme we present for stabilising the phase of multiple optical channels.  We utilise the reconfigurability of integrated photonic circuits to complete one iteration of phase stabilisation of any number of modes in just two rounds of measurement.
We carry this stabilisation out using photonic chips designed for the generation and characterisation of entangled qudits. Our method does not require any additional hardware to the quantum processes it enables.
To validate our approach when applied to the distribution of quantum states, we present experimental results on four-dimensional path-encoded entangled state distribution through noisy optical fibres with a quantum state fidelity of 86\%. 

The implementation of the phase stabilisation algorithm presented here is limited by the speed of electronics used to reconfigure the circuit and the need to interleave phase stabilisation measurements with the quantum measurement. Therefore, we expect significant improvements in measured fidelities and duty cycle if we utilise faster control electronics and phase modulators, in particular. The former could be achieved by a custom-designed multi-channel power supply. Regarding the latter, electro-optic phase modulators such as those based in  thin-film lithium niobate or barium titanate, have been shown to achieve on-chip speeds on the order of \SI{}{\giga\hertz} \cite{Chelladurai2025-BTOLiN, Psi2024-Hardware}. Integration of these to our system could be achieved by moving to a different material platform and/or heterogeneously integrating these high Pockels coefficient materials \cite{Psi2024-Hardware}. Improving the speed of inducing the phase shift even from \SI{}{\hertz} to \SI{}{\kilo\hertz} would improve the duty cycle from the current 18\% to at least 80$\%$. In addition, reducing the rate of phase drift in the channel by employing multi-core fibre instead of separate single-mode fibres in the inter-chip link would allow for the distribution of path-encoded states over metropolitan distances \cite{Bacco2021-MCFChar, DaLio2021-PathQKD2kmMCF,DaLio2020-2kmStableTransmission} (Supplementary Information).

With improvements to the speed of the hardware used for the phase stabilisation, in combination with taking steps to mitigate phase drift in the channel, this work demonstrates that path encoding is a feasible approach for the distribution of HD entangled quantum states between integrated quantum photonic processors at greater rates and over longer distances.

\section*{Funding and Acknowledgements}
M.A.T. wishes to thank E. Lavie for useful discussions, W.A. Murray for technical assistance, and J.F.F. Bulmer for receiver circuit design. M.A.T. highlights that results and methods presented here are present in the associated PhD thesis accessible at https://hdl.handle.net/1983/be0277fc-8107-4a1b-b4b5-4fc11df71dbd.
The authors acknowledge multiple funding sources, including: UK Quantum Technology Hub for Quantum Communications Technologies (EP/M013472/1), the QuantERA ERA-NET SQUARE project, and fellowship support from EPSRC (EP/N003470/1). D.B. acknowledges support from the European Union ERC StG, QOMUNE, 101077917. S.P. acknowledges funding from the VILLUM FONDEN research grant No. VIL60743, from the ERC Starting Grant ``ASPEQT'' (No. 101221875), and support from the NNF Quantum Computing Programme.

\section*{Author contributions}
J.W., M.B., and D.B. conceived and designed the experiment. 
Y.D. designed and fabricated the transmitter device.
M.A.T., D.L., S.P., and B.A.S. characterised the devices.
M.A.T., D.L., and B.A.S. built the experiment.
M.A.T., P.W.Y., D.L., and C.V. implemented the experiment.
M.A.T. performed the data analysis and
J.B., C.V., and P.W.Y contributed to the interpretation of results.
D.S., J.G.R., L.K.O., M.G.T., K.R., and J.B. managed the project.
M.A.T., P.W.Y., and J.B. wrote the manuscript with input from all the authors.

\newcommand{\ketbra}[2]{| #1 \rangle \langle #2 |}

\newcommand\SupplementaryMaterials{%
  \xdef\presupfigures{\arabic{figure}}
  \xdef\presupsections{\arabic{section}}
  \xdef\presupequations{\arabic{equation}}
  \renewcommand\thefigure{S\fpeval{\arabic{figure}-\presupfigures}}
  \renewcommand\thesection{S\fpeval{\arabic{section}-\presupsections}}
  \renewcommand\theequation{S\fpeval{\arabic{equation}-\presupequations}}
}

\newpage 
\clearpage
\onecolumngrid
 
\vspace{-0.2cm}

\SupplementaryMaterials

\begin{center}
\large\bfseries Supplementary Information
\end{center}

\section{Experimental set-up}

\begin{figure*}[b]
\centering
\includegraphics[width=\linewidth]{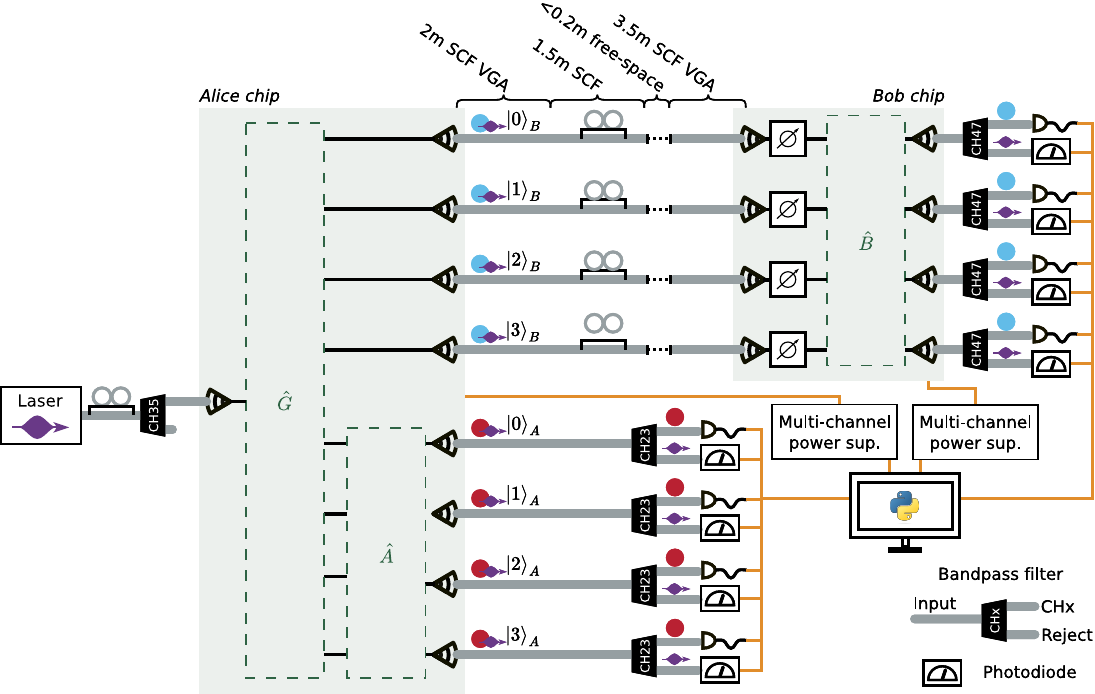}
\caption{Schematic of the experimental set-up including off-chip components.
}
\label{fig:setup_supps}
\end{figure*}

A more detailed schematic of the experimental set-up can be seen in Fig. \ref{fig:setup_supps}. Light from a continuous-wave laser (Yenista Optics Tunics-T100S-HP) is coupled into the Alice chip via one channel of a V-groove array (VGA) (Oz Optics). Prior to the VGA there is a polarisation controller (Newport in-line polarisation controller) to maximise coupling to a grating coupler, and a ITU channel 35 fibre-based bandpass filter (Fibre Stores FTTx Filter WDM) to remove sidebands from the laser. At the output of the Alice chip (coupled out via the same VGA), each signal path has a polarisation controller and free-space optical delay line (Oz Optics ODL-100). Each mode is then coupled into the Bob chip via another VGA. The phase shifters on the Alice and Bob chips are independently controlled using multi-channel power supplies (Qontrol Q8 modules) which are controlled via python scripting. All the signal and idler paths are VGA coupled out of their respective chips into two layers of channel 47 and channel 23 bandpass filters respectively to achieve \SI{60}{\decibel} of extinction between the single photons and residue pump and any photons generated in other resonances of the MRR. The pass channels containing the signal and idler photons are directed to single photon detectors (Photon Spot SNSPD system) where events are detected and recorded by a time-tagger (Swabian Instruments TimeTagger Ultra Performance). The reject channels of the filters are directed to photodiodes (Thorlabs PM100D) for classical characterisation of the system and phase stabilisation.

\section{Source characterisation}
\subsection{Source spectra}

\begin{figure*}
\centering
\includegraphics[scale=1]{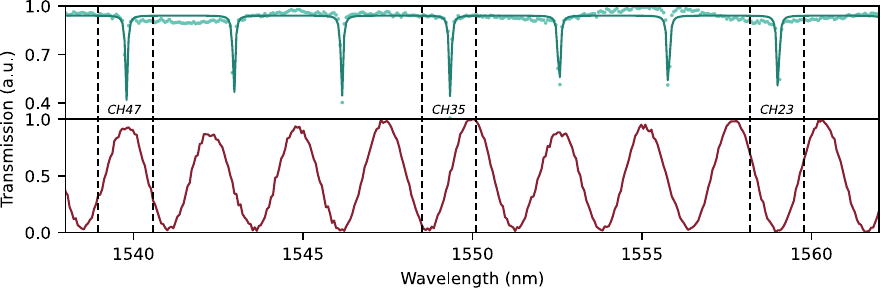}
\caption{MRR (top) and AMZI (bottom) spectrum.
}
\label{fig:AMZI_MRR}
\end{figure*}

The geometries of the MRR and AMZI components were designed such that the signal and idler wavelengths could be separated by the AMZI and that the signal, pump, and idler wavelengths align with ITU channels \cite{ITUGrid} such that standard telecommunications components could be used off-chip. The MRR has a diameter of \SI{55.36}{\micro\meter}, and consequently an FSR of 3.2nm. The path length difference between the two arms of the AMZI is \SI{217.372}{\micro\meter} and results in an FSR of \SI{2.56}{\nano\meter}. As such, when tuned with a TOPS to shift the spectrum, the MRR has resonances at, $\lambda_i=$\SI{1558.98}{\nano\meter}, $\lambda_p=$\SI{1549.30}{\nano\meter}, and $\lambda_s=$\SI{1539.77}{\nano\meter} (which correspond to ITU channels 23, 35, and 47) and the AMZI can be tuned with a TOPS such that $\lambda_i$ and $\lambda_s$ are emitted from different outputs of the AMZI - as can be seen from the spectra in Fig. \ref{fig:AMZI_MRR}.

\subsection{Source brightness}
The rate at which photon pairs are generated by each source affects the amplitude of that term in the quantum superposition we wish to prepare, and, for a fixed source design, can be tuned by changing the pump power, $P$.
For photons generated via SFWM, as in our SOI chip, the number of single signal photons ($S_s$), single idler photons ($S_i$), and coincidences ($C_{s,i}$) detected can be expressed as \cite{Bonneau2014-Thesis, Silverstone2016-SiliconQPhotonics, Llewellyn2019-Tele}:
\begin{equation}\label{eqn:source_char1}
    \begin{split}
    S_s & = \eta_s\gamma_{eff} P^2 +\beta_s P +B_{s},
    \\
    S_i & = \eta_i\gamma_{eff} P^2 +\beta_i P +B_{i},
    \\
    C_{s,i} & = \eta_s\eta_i\gamma_{eff} P^2 + B_{i,s},
    \end{split}
\end{equation} 
where $\eta_s$ and $\eta_i$ are the detection efficiencies of the signal and idler photons respectively; $\gamma_{eff}$ is the effective efficiency of the photon pair source; $\beta_s$ and $\beta_i$ are the linear noise contributions; and $B_s$, $B_i$, and $B_{s,i}$ are the background counts for the signal, idler, and accidental coincidences measured. Therefore, by measuring singles and coincidences of generated photon pairs and fitting the three quadratic equations of Eq. \ref{eqn:source_char1}, we can extract $\eta_s$, $\eta_i$, and $\gamma_{eff}$ to quantify photon pair generation and detection efficiencies. Fig. \ref{fig:source_char}a shows the relative brightness of our four MRR sources, which is used to inform the pump power we choose for each source in order to achieve a balanced superposition in the state preparation stage.

To increase state distribution rate we can increase the optical pump power, however, in silicon we are limited by non-linear effects such as two-photon absorption which limits pair production rate \cite{Husko2013-TPA}, and self-phase modulation which results in instability of the source \cite{Sinclair2016-SelfPhaseMod}, in addition to heating of the ring due to the laser causing instability \cite{Borghi2021-ThermalEffectOfPump, Novarese2022-ThermalEffectOfPump}. In addition, we wish to limit the probability of multi-pair production in the source, as such it is desirable to remain in a low-pumping regime.

\subsection{Aligning the source spectra}
\begin{figure*}
\centering
\includegraphics[scale=1]{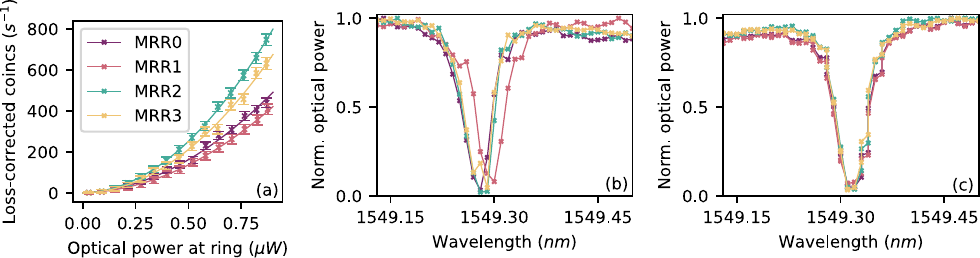}
\caption{(a) Source brightness characterisation. (b) Unaligned ring resonances. (c) Aligned ring resonances.
}
\label{fig:source_char}
\end{figure*}
Generating $d$-dimensional entangled states by coherently pumping $d$ photon
pair sources requires the sources to have a high degree of indistinguishability \cite{Wang2018-16D}. MRRs lend themselves well to this task, compared to basic waveguide sources in particular, as their resonant nature enhances spectral purity \cite{Burridge2020-MRRPurity}. However, due to fabrication imperfections, the spectrum of each MRR varies and the resonances of each source may not overlap spectrally, as can be seen in Fig. \ref{fig:source_char}b, impairing indistinguishability. Fortunately, due to TOPS on the MRRs we can tune the positions of the resonances in the spectrum of each ring in order to overlap them. Unfortunately, however, thermal cross-talk between components occurs meaning that when the spectrum of one ring is adjusted, it will affect the spectra of others. As such, we maximise spectral overlap of MRRs using the following method:
\begin{enumerate}
    \item Direct the amount of pump power to all sources as will be used in the measurement - changing pump power can change the properties of a MRR due to the aforementioned cross- and self-phase modulation and heating effects, hence affecting resonance position.
    \item Perform an initial scan of the spectrum of each individual MRR, $n$, by scanning the voltage of the TOPS, $V_n$, and measuring the optical power of the pump wavelength, $P_n(\phi_n)$. This allows us to find an initial value for $V_n$ corresponding to the MRR being on-resonance with the pump - this will be where minimum power is measured.
    \item Set each ring to the initial voltages found in the previous step.
    \item Iteratively adjust each $V_n$ by small amounts while measuring the optical power of each MRR in order to minimise $\sum_{n}P_n(V_n)$. This results in the minimum pump power being detected after the rings, and therefore maximum coupling of the pump wavelength into the ring, serving to maximise the overlap of the pump resonance of the MRRs.
\end{enumerate}
The resulting overlapping spectra of our sources can be seen in Fig. \ref{fig:source_char}c.

\subsection{Characterising indistinguishability}
\begin{figure*}
\centering
\includegraphics[scale=1]{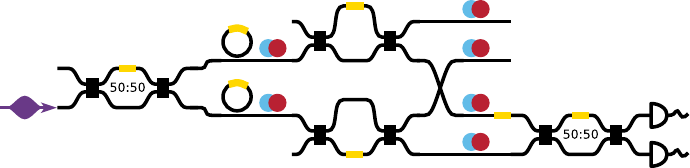}
\caption{Schematic of the circuit used to perform RHOM measurements between a pair of sources.
}
\label{fig:RHOM}
\end{figure*}

Following spectral alignment of the sources, we characterise their indistinguishability. The Hong-Ou-Mandel (HOM) effect describes the effect of a 50:50 beamsplitter on the Fock state 
\begin{equation}
    \ket{11} = \hat{a}^\dagger_{0}\hat{a}^\dagger_{1}\ket{00}.
\end{equation}
Here, the first index of the state corresponds to the first path mode and the second corresponds to the second path mode, with $\hat{a}^\dagger_{0}$ and $\hat{a}^\dagger_{1}$ being the creation operators in those modes respectively. 
The effect of a 50:50 beamsplitter on light input to the two modes is given by the transformations
\begin{equation}
    \begin{split}
        \hat{a}^\dagger_0\rightarrow \frac{1}{\sqrt{2}}(\hat{a}^\dagger_0+\hat{a}^\dagger_1)
        \\
        \hat{a}^\dagger_1 \rightarrow  \frac{1}{\sqrt{2}}(\hat{a}^\dagger_0-\hat{a}^\dagger_1),
    \end{split}
\end{equation}
where $\hat{a}^\dagger_{0}$ and $\hat{a}^\dagger_{1}$ then correspond to the creation operators of the first and second outputs of the beamsplitter. As such, the output when the state $\ket{11}$ has passed through the beamsplitter is 
\begin{equation}
    \ket{\psi}_{HOM}=
    \frac{1}{2}((\hat{a}^\dagger_{0}+ \hat{a}^\dagger_{1}) (\hat{a}^\dagger_{0}-\hat{a}^\dagger_{1}))\ket{00}
    \\
    =\frac{1}{2}(\hat{a}^\dagger_{0}\hat{a}^\dagger_{0}+\hat{a}^\dagger_{1}\hat{a}^\dagger_{0}-\hat{a}^\dagger_{0}\hat{a}^\dagger_{1}+\hat{a}^\dagger_{1}\hat{a}^\dagger_{1})\ket{00}.
\end{equation}

When the photons are indistinguishable we see no coincidences as $\hat{a}^\dagger_{1}\hat{a}^\dagger_{0}=\hat{a}^\dagger_{0}\hat{a}^\dagger_{1}$, and the final state is
\begin{equation}
    \ket{\psi}_{HOM}=\frac{1}{\sqrt{2}}(\hat{a}^\dagger_{0}\hat{a}^\dagger_{0} + \hat{a}^\dagger_{1}\hat{a}^\dagger_{1})\ket{00}=\frac{1}{\sqrt{2}}(\ket{20}+\ket{02}).
    \label{eqn:HOM_output}
\end{equation}

In fact, when pumping two photon pair sources such that there is an equal probability of either generating a pair of photons at once (and post-selecting on only one source firing at once), the prepared state in the Fock basis is Eq. \ref{eqn:HOM_output}. As such we can look at the `time-reversed' HOM (RHOM) effect \cite{Silverstone2014-OnChipInterference} where the initial state is 
\begin{equation}
    \frac{1}{\sqrt{2}}(\hat{a}^\dagger_{0}\hat{a}^\dagger_{0} + e^{i\phi}\hat{a}^\dagger_{1}e^{i\phi}\hat{a}^\dagger_{1})\ket{00},
\end{equation}
which is the same as the output Eq. \ref{eqn:HOM_output} but now we introduce a relative phase, $\phi$, between the sources. When this state is incident on a beamsplitter, the output state is
\begin{equation}
    \ket{\psi}_{RHOM} = \frac{1}{2\sqrt{2}}((\hat{a}^\dagger_{0}+\hat{a}^\dagger_{1})(\hat{a}^\dagger_{0}+\hat{a}^\dagger_{1}) + e^{i2\phi}(\hat{a}^\dagger_{0}-\hat{a}^\dagger_{1})(\hat{a}^\dagger_{0}-\hat{a}^\dagger_{1}))\ket{00},
    \label{eqn:RHOM_2phi}
\end{equation}
which can be represented as
\begin{equation}
    \ket{\psi}_{RHOM} = \frac{1}{2\sqrt{2}}((\hat{a}^\dagger_{0}\hat{a}^\dagger_{0} + \hat{a}^\dagger_{1}\hat{a}^\dagger_{1})\cos(\phi) + i(\hat{a}^\dagger_{0}\hat{a}^\dagger_{1} + \hat{a}^\dagger_{1}\hat{a}^\dagger_{0})\sin(\phi))\ket{00}.
    \label{eqn:RHOM_fringe}
\end{equation}
From this we can see the cos term corresponds to both photons at the same output (therefore giving no coincidences) and the sin term corresponds to the photons at separate outputs (resulting in a coincidence). As such, by varying phase we can observe a fringe in coincidence counts between the two outputs where any non-zero value at the minimum of the fringe comes from contributions from distinguishable photons which reduces the fringe visibility defined as
\begin{equation}
    \mathcal{V} = \frac{I_{max}-I_{min}}{I_{max}+I_{min}},
    \label{eqn:visibility} 
\end{equation}
where $I_{max}$ and $I_{min}$ are the maximum and minimum intensities of the fringe respectively. This allows us to quantify the indistinguishability of two photon pair sources, with the results presented in Fig. \ref{fig:setup}c. A schematic for the subset of the Alice chip used for this measurement can be seen in Fig. \ref{fig:RHOM}. These measurements are conducted on one chip (instead of sending the signal photons to the Bob chip) and the coincidence measurements are background-subtracted.  When classical light is input to the same system, a fringe in optical power with half the frequency of the quantum fringe is observed, as shown in Fig. \ref{fig:setup}b, highlighting a difference between the classical and quantum phenomena.

\section{Inter-chip path length matching}
The four path modes connecting Alice and Bob must be the same length in order for the modes to temporally overlap at the interferometer on Bob \cite{Fox2006-QuantumOptics} and to avoid wavelength-dependent phase effects in the inter-chip link (i.e. to prevent an inter-chip AMZI being formed instead of an MZI).
The latter is particularly relevant where the phase stabilisation is concerned as we use pump wavelength light to correct for phase drift of the signal wavelength photons.
The constraint on temporal alignment imposed by the phase stabilisation requires sub-picosecond precision: the separation between the two wavelengths used is \SI{9.53}{\nano\meter} meaning a $\pi$ phase discrepancy between the wavelengths occurs with \SI{0.83}{\pico\second} of path length mismatch in free-space.

We achieve coarse path length matching by ensuring each path between Alice and Bob used approximately the same length of optical fibre, and we achieved fine adjustment using a tunable free-space delay line. In order to set the tunable delay lines to path-length match the modes with different levels of precision, three methods were used, which will be outlined in the following, with increasing precision capability. In all methods outlined here, we begin by setting up a MZI across the system such that the two arms of the MZI constitute the paths we wish to length-match - in reality, this forms an AMZI if the path lengths are not matched. To length-match all four modes in our system we rely on the reconfigurability of the integrated photonic chips and assume on-chip path length matching.

\subsection{Coincidence measurement of pulsed laser}
\begin{figure*}
\centering
\includegraphics[scale=1]{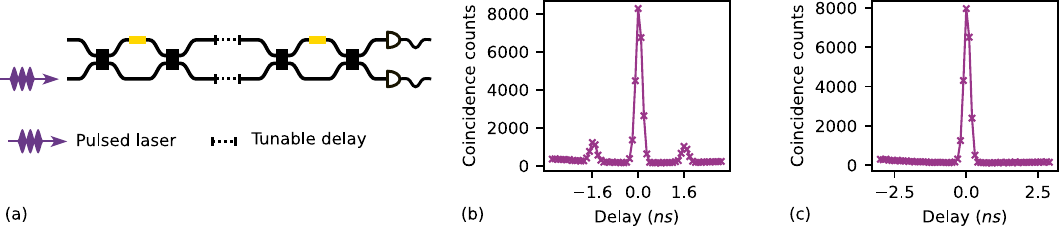}
\caption{(a) Schematic for path-length matching through measurement of a coincidence histogram with a pulsed laser. (b) 
Coincidence histogram when paths are not length-matched. (c) Coincidence histogram when paths are length-matched.}
\label{fig:delay1}
\end{figure*}

To evaluate time delays on the order of hundreds of picoseconds, we use a pulsed laser (PriTel Inc. Femtosecond fibre laser) as the light source, leveraging its short coherence time, and heavily attenuate it such that it can be detected by single photon detectors. The initial state when we consider just two photons from a single pulse and input those to mode 0 of the MZI is
\begin{equation}
    \hat{a}^\dagger_0\hat{a}^\dagger_0\ket{00} = \ket{20}.
\end{equation}
When this passes through the first 50:50 beamsplitter of the MZI the output is
\begin{equation}
    \frac{1}{2}(\hat{a}^\dagger_0\hat{a}^\dagger_0 + \hat{a}^\dagger_1\hat{a}^\dagger_0 + \hat{a}^\dagger_0\hat{a}^\dagger_1 + \hat{a}^\dagger_1\hat{a}^\dagger_1)\ket{00}.
\end{equation}
At this point the state passes through the channel with mismatched path lengths. As just the relative path length difference is relevant, we will say there is a time difference of $\tau$ on the second mode and indicate this with a prime on the creation operator. The state after passing through the mismatched paths is then
\begin{equation}
    \frac{1}{2}(\hat{a}^\dagger_0\hat{a}^\dagger_0 + \hat{a'}^{\dagger}_1\hat{a}^\dagger_0 + \hat{a}^\dagger_0\hat{a'}^\dagger_1 + \hat{a'}^\dagger_1\hat{a'}^\dagger_1)\ket{00}.
\end{equation}
Finally, following the second beamsplitter of the MZI the state which is passed to the detectors is
\begin{equation}
    \frac{1}{2\sqrt{2}}(
    \hat{a}^\dagger_0\hat{a}^\dagger_1 + 
    \hat{a}^\dagger_1\hat{a}^\dagger_0 -
    \hat{a'}^{\dagger}_0\hat{a'}^\dagger_1 - 
    \hat{a'}^{\dagger}_1\hat{a'}^\dagger_0 +
    \hat{a'}^\dagger_0\hat{a}^\dagger_1 +
    \hat{a}^\dagger_1\hat{a'}^\dagger_0 - 
    \hat{a'}^\dagger_1\hat{a}^\dagger_0 - 
    \hat{a}^\dagger_0\hat{a'}^\dagger_1)\ket{00},
\end{equation}
where we have disregarded any terms that do not result in a coincidence event at the detectors and renormalised the state. The first four terms correspond to coincidence events where there is no relative time delay between the modes, resulting in a peak at $t=0$. The next two terms correspond to coincidences where photons at the first detector arrive $\tau$ seconds after photons at the second detector, resulting in a peak at time $t=-\tau$. The final two terms correspond to photons arriving $\tau$ seconds after those at the first detector, resulting in a coincidence peak at time $t=\tau$. As such, we can determine the value of $\tau$, corresponding to the path length difference based on the separation of the peaks in a coincidence histogram. This allows us to align path lengths to a precision on the scale of the detector jitter which, here, is on the order of \SI{100}{\pico\second}. A schematic for implementing this method can be seen in Fig. \ref{fig:delay1}a. Coincidence histograms of mis-aligned and coarsely aligned delay lines are shown in Fig. \ref{fig:delay1}b and c respectively.

\subsection{Spectral oscillations in broadband source}
\begin{figure*}
\centering
\includegraphics[scale=1]{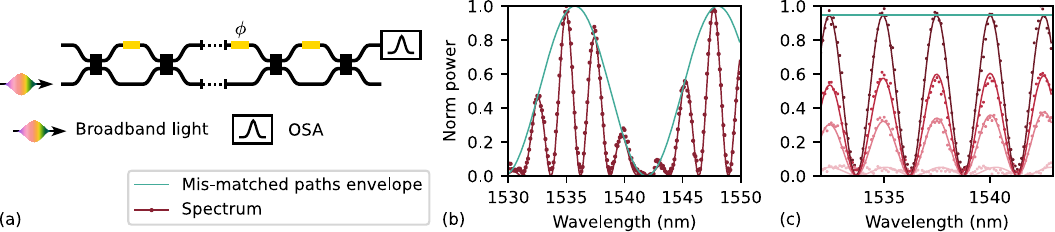}
\caption{
(a) Schematic for path-length matching using the wavelength-dependent effect of the paths not being length-matched. (b) 
Output of the interferometer when the paths are not length-matched. (c) Output of the interferometer when the paths are length-matched. The different fringes are the output spectrum at different phases $\phi$.}
\label{fig:delay2}
\end{figure*}
In this method, again we set-up the same interferometer as before but instead use a broadband classical light source (Yenista optics OSICS SLD 1550nm) and measure the spectral output using an optical spectrum analyser (OSA) (Anritsu MS9740A OSA) in order to capture the wavelength-dependent output, as shown in Fig. \ref{fig:delay2}a. If we observe a fringe in the spectrum as measured on the OSA then this indicates the interferometer acts differently on different wavelengths due to path-length mismatch, as shown in \ref{fig:delay2}b. However, if no fringe is present and, when we change a phase shifter on one of the arms, the power changes uniformly across the spectrum, this indicates the paths are balanced. If no change in the spectrum is observed as the phase shifter is swept then this indicates the path lengths are misaligned greater than the coherence of the photons, as shown in \ref{fig:delay2}c. This method allows for alignment with a precision dependent on the bandwidth of the broadband source and the OSA measurement, $\sim$\SI{10}{\nano\meter} in our implementation. This corresponds to a free-space mismatch of \SI{0.4}{\pico\second}, corresponding to alignment within $\pi/2$ for our wavelengths. The \SI{2.56}{\nano\meter} period fringes present in both Fig. \ref{fig:delay3}b and c are due to the on-chip AMZIs which are located on the paths we are length-matching.

\subsection{Two-wavelength phase fringe}
\begin{figure*}
\centering
\includegraphics[scale=1]{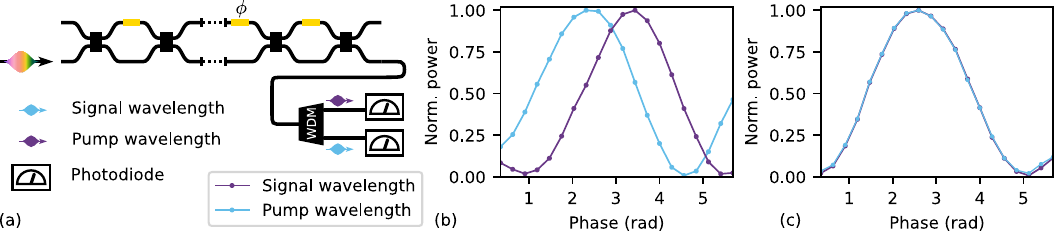}
\caption{(a) Schematic for path-length matching looking at the signal and pump wavelengths at the output. (b) Optical power fringes when the paths are not length-matched. (c) Optical power fringes when the paths are length-matched.
}
\label{fig:delay3}
\end{figure*}
Here we use the same set-up as the previous method but replace the OSA with a wavelength-division multiplexer and two photodiodes (as shown in Fig. \ref{fig:delay3}a), allowing us to look at the effect of the inter-chip interferometer on the signal and pump wavelengths specifically. The output power of the interferometer with path length mismatch in free-space $\Delta L$ for wavelength $\lambda$ is
\begin{equation}
    P(\lambda, \Delta L, \phi) \propto \sin^2\left(\frac{\pi \Delta L}{\lambda}+\frac{\phi}{2}\right).
\end{equation}
This allows us to extract $\Delta L$ from the difference in phase offset between the two wavelength fringes measured by changing $\phi$. Measured fringes in signal and pump power can be seen in Fig. \ref{fig:delay3}b and c when they are mis-aligned and aligned respectively.

\section{Loss balancing and correction}
Optical loss affects the system in two particular ways: overall loss affects state distribution rate, while relative losses between the paths affect the fidelity of state distribution and measurement.
The main causes of loss in the system are when light is coupled on- and off-chip, and when light propagates through the fibre links. The signal photons are affected by relative losses in the inter-chip link before the projectors are applied. Both the signal and idler photons are affected by relative losses after the projectors are applied and the photons are sent to the detectors. While the latter can be corrected in post-processing of the measured statistics, the former must be corrected in the physical system as they affect the fidelities of the operators we apply. We address both points of loss in the following.

\subsection{Loss balancing between the chips}

To mitigate the effect of relative losses between different paths for the signal photons in the inter-chip link, we characterise and correct for them in the physical system. We measure the relative losses on a $d$-mode system by configuring the Alice interferometer to direct light with power $P$ to path $x$ which is then coupled off-chip, sent though the inter-chip link, and coupled into the $x$-th mode of the Bob chip which has the interferometer $\hat{B}$ set to direct light to output $y$. The optical power, $O_{x,y}$, measured at the output can be expressed as \begin{equation}
    O_{x,y} = \alpha_x\eta_y P \quad x,y \in[0,d-1].
\end{equation} 
where $\alpha_x$ is the efficiency of the channel up to and including the coupling into the Bob chip, and $\eta_y$ is the combined coupler, channel, and photodiode efficiency.

In order to balance the optical channel efficiencies, $\alpha_x$, between the modes, we include variable attenuators on each mode such that losses of each mode can be matched to that of the highest loss mode. As such, we are only interested in the relative efficiencies between two modes $x_0$ and $x_1$, $\alpha_{x_0}/\alpha_{x_1}$, and not the absolute efficiencies. This means the relative efficiencies, independent of $\eta_y$, are
\begin{equation}
    \frac{\alpha_{x_0}}{\alpha_{x_1}} = \frac{\alpha_{x_0}\gamma_{x_0} P}{\alpha_{x_1}\gamma_{x_0} P} = \frac{O_{x_0,x_0}}{O_{x_1,x_0}}
\end{equation}
and can be obtained from measurements of output optical power for different I/O combinations by reconfiguring the Alice and Bob interferometers, assuming on-chip losses of each I/O path through $\hat{B}$ to be even. The lossiest mode ($x=l$) is the mode for which 
\begin{equation}
    \frac{\eta_{l}}{\alpha_{x}}<1 \quad \textrm{for all }x\neq l
\end{equation}
and $1-\eta_{l}/\alpha_{x}$ is the amount mode $x$ must be attenuated by in order to match the channel efficiency of mode $l$. We implement the loss balancing using MZIs on each path mode at the input of the Bob chip where one output of each MZI is simply a reject port. The phase that must be set in the MZI attenuator on mode $x$ is given by
\begin{equation}
    \phi_x = 2\sin^{-1}\left(\sqrt{\frac{\eta_l}{\eta_x}}\right)
\end{equation}
in order to achieve chip-to-chip loss balancing between the modes.

\subsection{Loss correction after the chips} \label{sec:dce}

We construct the density matrix of a quantum state by measuring many coincidence detection events between pairs of detectors $a$ and $b$ following state projection with $\hat{A}$ and $\hat{B}$. We then normalise the coincidence events, $C_{a,b}$, by $\sum_{a,b}C_{a,b}$ to give the probability, $P_{a,b}$ associated with that outcome:
\begin{equation}
    P_{a,b} = \frac{\eta_{a}\eta_{b} C_{a,b}}{\sum_{a,b}C_{a,b}},
    \label{eqn:dce}
\end{equation}
where $\eta_{a}$ and $\eta_{b}$ are the detector channel efficiencies of the two detectors in the coincidence measurement, and $\eta_{a}\eta_{b} C_{a,b}$ is the number of \textit{detected} coincidences between $a$ and $b$. To correctly calculate $P_{a,b}$, we must characterise $\eta_{a,x}$ and $\eta_{b,y}$. In order to do this, we generate signal and idler photon pairs from one photon source and direct the signal (idler) photons to each Bob (Alice) detector in turn and measure the number of detected photons.
The efficiency values need only be relative due to the normalisation in Eq. \ref{eqn:dce}, therefore we define the detector with the highest photon counts as 100\% efficient and define all other detector efficiencies relative to this.
It should be noted that it is essential that losses before the outputs of $\hat{A}$ and $\hat{B}$ are balanced, as such we first implement the inter-chip loss balancing described above.

\section{Phase stabilisation}

While the phase stabilisation algorithm presented is simple in principle - measuring MZI fringes where the phase offset corresponds to the phase drift of the paths in the interferometer - there are important considerations that make this method possible. 

Firstly, we must consider the number of points to measure on the fringe. As each point requires a finite time to measure, and we wish to minimise the time taken for phase stabilisation, it is desirable to measure the minimum number of points while maintaining accuracy of the fit to the fringe. In order to carry out the fitting of Eq. \ref{eq:fringe} to the measured fringe, we used the non-linear least-squared Python function scipy.optimize.curve\_fit. This requires the number of data points to be greater than or equal to the number of degrees of freedom in the fitting function - in our case this number is four: $E_{0,n}$, $P_{max}$, $P_{min}$ and $\nu$. It was found that, given reasonable spacing between the points, parameters bounds, and parameter initial guesses, four points provides sufficient fitting accuracy.

Reasonable bounding of the parameters and initial guesses are obtained by adding a characterisation step to the process, whereby the system is set up exactly as it will be in the experiment and phase stabilisation fringes are measured, except with a greater number of points measured on the fringe. In this context we can afford the extra cost of time imposed by measuring more points as we are not using these fringes to actively stabilise the system. The greater number of points allow us to accurately characterise the mean, $\overline{x}$, and standard deviation, $\sigma_{x}$, of the $P_{max}$, $P_{min}$ and $\nu$ parameters from just 30 fringes per configuration of the phase stabilisation.
Using $\overline{x}$ as the initial guesses and $\overline{P}_{max}\pm3\sigma_{P_{max}}$, $\overline{P}_{min}\pm2\sigma_{P_{min}}$, and $\overline{\nu}\pm2\sigma_{\nu}$ as the bounds enables accurate fitting to the four-point fringes in the active phase stabilisation.

\begin{figure*}
\centering
\includegraphics[scale=1]{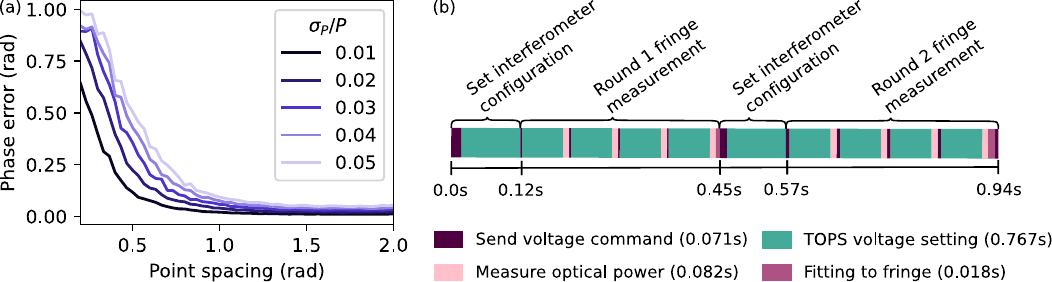}
\caption{
(a) Phase error resulting from fitting to simulated fringes with different point spacings for different levels of optical power noise. (b)  Time accumulated for each stage of one iteration of the phase stabilisation protocol as-implemented.}
\label{fig:PS_char_supps}
\end{figure*}

Fig. \ref{fig:PS_char_supps}a shows the error in phase offset value obtained from fitting Eq. \ref{eq:fringe} to simulated noisy fringes with four points with different spacing between them. We simulated for different levels of noise (quantified as standard deviation divided by power, $\sigma_P/P$) based on typical data measured from the real system and in all cases, phase error was found to plateau after 1.4rad.

\subsection{Timing}
\begin{figure*}
\centering
\includegraphics[scale=1]{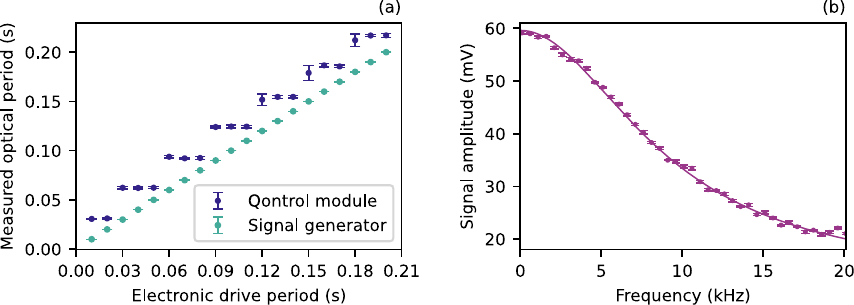}
\caption{
(a) Measured period of change in optical power output when compared to the period of the driving electronic signal. (b) Optical response of a TOPS as drive frequency is varied.
}
\label{fig:speed_char}
\end{figure*}
A timeline of one iteration of the phase stabilisation can be seen in Fig. \ref{fig:PS_char_supps}b. 82\% of one iteration is taken by the time it takes to set a configuration of the circuit. To characterise this further, we look at the two components involved in this step: the TOPS and the electronics to control it. For the latter, we drive a MZI TOPS with a channel of the multi-channel power supplies (Qontrol Q8 modules) used in the experiment and compare to driving the TOPS with a signal generator. We can see from Fig. \ref{fig:speed_char}a, that it takes \SI{0.03}{\second} for the Qontrol modules to apply a voltage to a TOPS, this is due to the clock cycle of the electronics. For the former, we drive a MZI TOPS with a signal generator and observe the amplitude of the resulting optical modulation. From Fig. \ref{fig:speed_char}b, we can see the phase shifter has a \SI{3}{\decibel} bandwidth of \SI{11.7}{\kilo\hertz}, however to achieve 100$\%$ signal amplitude, the TOPS is limited to $\sim$\SI{1}{\kilo\hertz}. This is in line with typical SOI-based TOPS \cite{Sun2022-SiPhaseShifters}. This means that here we are limited by the speed of the electronics, not the TOPS. While the benefit of the electronics we used are their multi-channel capability, technology is available for multichannel control at higher speeds, meaning it is reasonable to conclude faster phase stabilisation can be implemented with updated electronics to control the TOPSs.

\subsection{Error scaling with dimension}
\begin{figure*}
\centering
\includegraphics[scale=1]{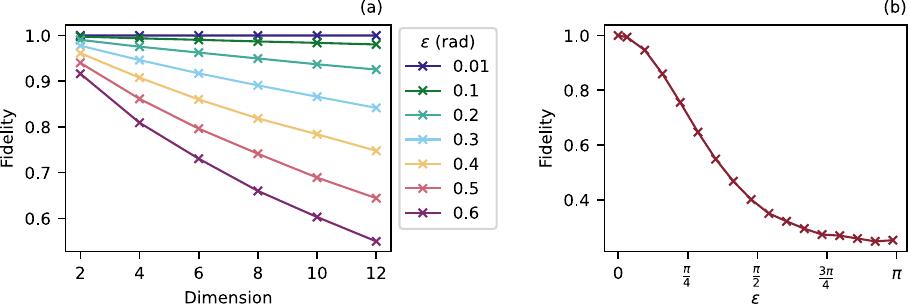}
\caption{(a) Simulated fidelity of states of different dimension for different $\epsilon$. (b) Simulated fidelity of $d=4$ state distribution dependent on $\epsilon$.
}
\label{fig:scaling_of_method}
\end{figure*}

As presented, the phase difference between every mode, $n$, and the reference mode can always be obtained from just two rounds of measurement for a system of any dimension by interfering modes $n$ and $n+1$ in the first round and $n+1$ and $n+2$ in the second round. This allows $\Delta_{0,n}=\Delta_{0,n-1}+\Delta_{n-1,n}=\Delta_{0,n}=\Delta_{0,n+1}+\Delta_{n+1,n}$ to be inferred for any phase difference not measured directly against the reference mode. Given an average error $\epsilon$ on a direct phase measurement, this means that the maximum error on an inferred phase difference is $\epsilon d/2$.
Considering this, we simulate how effectively we can stabilise a system with this method for increasing $d$.
In the simulation, for each ${n,n'}$ pair, a normal distribution determined by $\epsilon$ was sampled from to obtain erroneous $\Delta\phi_{n,n'}$ values which are used to obtain erroneous $\Delta\phi_{0,n}$ values for all $n$. We the calculate the fidelity of the simulated outcome compared to the ideal when the state is measured in the Hadamard basis.
Fig. \ref{fig:scaling_of_method}a shows simulated fidelity for even dimensions up to 12 for different levels of error. Fig. \ref{fig:scaling_of_method}b looks specifically at how error in phase measurement affects the $d=4$ case, indicating that with our 95\% measured fidelity from Fig. \ref{fig:phase_stab}c, we have an average phase measurement error of 0.2rad.

\section{2D state distribution}
\begin{figure*}
\centering
\includegraphics[scale=1]{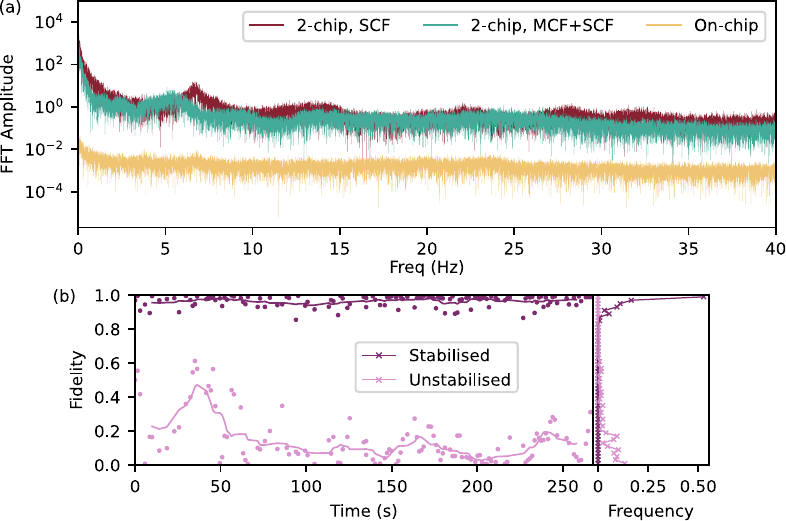}
\caption{
(a) Frequency spectrum of the power drift of each two-mode interferometer. (b) Fidelity of the classical Hadamard basis state measurement over time without and with phase stabilisation with the SCF+MCF inter-chip link.}
\label{fig:2D_PS_char}
\end{figure*}

As a prerequisite to the distribution of high-dimensional states, here we assess phase drift in a two-mode interferometer and look at the distribution of a 2D Bell state.
We show three configurations of our system: a fully on-chip interferometer; a two-chip interferometer where the inter-chip link consists of single-core fibre (SCF), as used in the experiment; and a two-chip interferometer where part of the SCF inter-chip link is replaced with multicore fibre (MCF). MCF consists of multiple cores that allow multiple path modes to be contained in the same fibre. MCF has been shown to enable the distribution of path encoded states with greater stability as each mode experiences similar effects from the environment due to being in the fibre \cite{Bacco2021-MCFChar, DaLio2020-2kmStableTransmission}.

Fig. \ref{fig:2D_PS_char}a shows a FFT of the optical power measured at the output of the 2-mode interferometers as phase is allowed to drift in these three configurations. When comparing the \SI{7}{m} SCF link to the \SI{4}{m} SCF + \SI{3}{m} MCF, we can already see the benefit of even part of the SCF with MCF as phase drift is reduced, particularly the most significant contributions, and the rate of the most significant phase drift is reduced. This is also evidenced by Fig. \ref{fig:2D_PS_char}b where we can see the fidelity of the 4D $H_4^+$ classical state distributed through the SCF+MCF link, which can be compared to the full SCF link shown in Fig. \ref{fig:phase_stab}. In the MCF+SCF link the stabilised average fidelity in this sample is 97\% compared to 95\% in the SCF link.
Here we note, in this implementation, the \SI{4}{m} of SCF are present due to: the arrangement of grating couplers on the Alice chip being designed for a V-groove array, consisting of SCF; and polarisation controllers and delay lines still being a requirement to act on each path mode individually. However, the SCF in this link could be reasonably reduced to less than \SI{1}{m} by efficiently splicing the fibre components together and/or swapping the coupling approach to Alice from a VGA to MCF. This would significantly reduce the phase drift in the link as much of the contribution is from the SCF portion of the link, which would allow for distribution of path-encoded quantum states over km of MCF as shown in \cite{DaLio2021-PathQKD2kmMCF,Bacco2021-MCFChar}.

\begin{figure*}
\centering
\includegraphics[width=\linewidth]{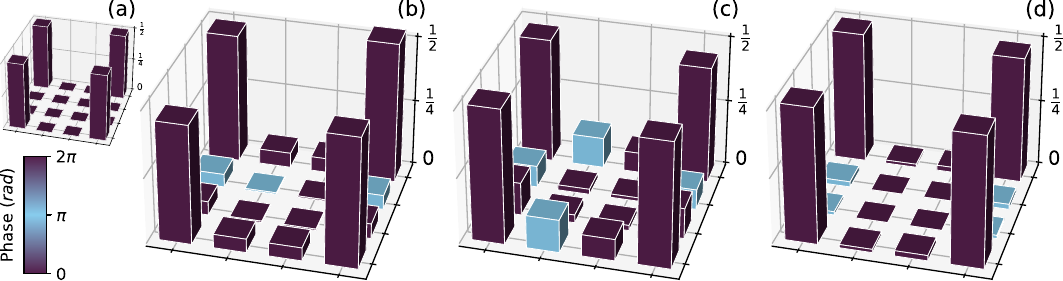}
\caption{
(a) Ideal density matrix of the 2D Bell-state $\ket{\Phi^+_2}$. Reconstructed density matrices for the experimentally prepared state (b) probed entirely on Alice, (c) with the signal photons sent to Bob through the SCF link and (d) with the signal photons sent through a SCF+MCF link.}
\label{fig:2D_tomo}
\end{figure*}

Fig. \ref{fig:2D_tomo} shows the reconstructed density matrices for the 2D Bell state $\ket{\Phi^+_2}=(\ket{00}+\ket{11})/\sqrt{2}$ (a) in the ideal case, (b) probed entirely on Alice, (c) with the signal photons sent to Bob through the SCF link, and (d) with the signal photons sent through the SCF+MCF link. In the first case, the fidelity of the measured state when compared to the ideal state is $98.7\pm0.3\%$ and its entanglement entropy $0.995\pm0.001$. In the second case, the fidelity of the measured state is $95.9\pm0.3\%$ with an entanglement entropy of $0.995\pm0.001$. The drop in fidelity between the two-chip and one-chip measurements can be attributed largely to imperfect phase stabilisation. In spite of this, the entanglement entropy remains the same, indicating no reduction of quantum correlation between the photons. In the third case, the fidelity of the measured state is 99.8$\pm0.02\%$, showing an improvement of $3.9\%$ from the tomography performed with the fully SCF inter-chip link and an improvement of $1.1\%$ fidelity over even the fully on-chip measurement can also be seen. This can be attributed to thermal cross-talk that occurs in the one-chip measurements as the projection interferometer on Alice used for tomography must be changed after the MRR resonances are aligned, causing them to shift. This cross-talk does not occur in the two-chip measurements as the Alice circuit need not be changed after the MRR resonances are aligned in this case.

\section{Quantum state tomography}\label{sec:QST}

Here we aim to provide a practical guide to performing quantum state tomography (QST), building up to the tomography of multiple, high-dimensional particles, as presented in this work. The density matrix of any state $\rho$ can be expressed as a linear combination of states $\ket{e_i}\bra{e_i}$ that span the Hilbert space, weighted by the probability $P_i$ of $\rho$ being in the state $\ket{e_i}\bra{e_i}$ \cite{Nielsen_Chuang}:
\begin{equation}
    \rho = \sum_{i}P_i\ket{e_i}\bra{e_i}.
\end{equation}
Mathematically, there are a number of key properties of a physical density matrix:
\begin{itemize}
    \item $\rho$ is a positive semi-definite operator. This means that all probabilities are positive or zero.
    \item $tr(\rho)=1$ which captures that all probabilities sum to 1.
    \item $\rho$ is self-adjoint ($\rho=\rho^\dagger$) meaning it is a Hermitian operator.
    \item $\rho$ being positive semi-definite and Hermitian also means all eigenvalues are real, as required for physical measurement outcomes.
\end{itemize}

In general, we wish to validate our experimentally prepared state $\tilde{\rho}$ against the ideal state $\rho$.
The characteristics of $\tilde{\rho}$ that we are particularly interested in are: entanglement, dimensionality, and fidelity with respect to the ideal state $\rho$.
If we are able to obtain the density matrix $\tilde{\rho}$ then we can extract these values. Experimentally, we do this via ``quantum state tomography'' (QST).

Tomography is the process of building a representation of something by performing many measurements. Classical tomography of a subject can be performed by taking multiple different measurements of the same system. However, upon measurement of a quantum system its state is perturbed, therefore multiple measurements on the same subject cannot be used to build a tomography. In QST, therefore, we prepare many copies of the same state such that we can carry out these measurements multiple times to build a tomography of the density matrix \cite{Altepeter2005-QST}.

These different measurements are performed by projecting on to the eigenstates $\ket{e_{i,k}}$ of a set of measurement operators $\hat{\sigma}_i$ that span the Hilbert space such that we measure the probabilities $P_{i,k}=Tr(\ket{e_{i,k}}\bra{e_{i,k}}\tilde{\rho})$, which corresponds to the probability of $\tilde{\rho}$ being in the state $\ket{e_{i,k}}\bra{e_{i,k}}$ \cite{Altepeter2005-QST}. Since we need to perform these projective measurements on many copies of $\tilde{\rho}$ to build the tomography, it is important to keep in mind how many measurements it is experimentally feasible to perform. Ideally we would like to use the minimum number of projections required to accurately construct $\tilde{\rho}$. Quantum states with $N$ qudits of dimension $d$ are defined by $d^{2N}$ parameters \footnote{A $D\times D$ matrix has $D^2$ elements. As a density matrix is Hermitian, the $D$ diagonal elements are real, requiring $D$ parameters to describe them. The $D^2-D$ off-diagonal elements are complex, requiring $2(D^2-D)$ parameters to describe them. However, the Hermitian properties also mean, the off-diagonal elements are the complex conjugate of those on the opposite side of the diagonal, therefore actually only $D^2-D$ parameters are required for the off-diagonals. Therefore a total of $(D)+(D^2-D)=D^2$ parameters are needed to describe a density matrix, where $D=d^{2N}$.}
therefore, we will need to obtain at least $d^{2N}$ measurement outcomes to characterise the state. The exact measurements required and how to construct the density matrix from them for different states is considered in the upcoming Section \ref{linear state recon}.

In the following we will outline the construction of density matrices of increasing size, starting from a single qubit ($d=2$, $N=1$), then n qubits ($d=2$, $N=N$), followed by a single qudit ($d=d$, $N=1$) and finally the most general case of multiple qudits ($d=d$, $N=N$). This will lead us to how we experimentally perform QST on our $d=4$, $N=2$ quantum state using separable projective measurements.

    \subsection{Linear state reconstruction}\label{linear state recon}
    \subsubsection{Single qubit}
    The essence of performing QST is that we wish to construct a matrix that represents a state. Any $D\times D$-sized matrix $\rho$ can be expressed as a linear combination of $D^2$ matrices that span the Hilbert space. For example for $D=2$, we could construct any $2\times 2$ matrix with the combination
    \begin{equation}
        \rho=
        S_0\begin{pmatrix}
            1&0\\
            0&0
        \end{pmatrix}+
        S_1\begin{pmatrix}
            0&0\\
            0&1
        \end{pmatrix}+
        S_2\begin{pmatrix}
            0&1\\
            0&0
        \end{pmatrix}+
        S_3\begin{pmatrix}
            0&0\\
            1&0
        \end{pmatrix},
    \end{equation}
    where $S_i$ are complex coefficients that weight the contributions of each matrix. To perform tomography we wish to obtain the coefficients from physical measurement of the expectation values of the matrices. For this, all operator matrices must be Hermitian, meaning measurements correspond to physical observables which, unfortunately, the simple construction above does not abide. A common set of operators, $\hat{\sigma}_i$, for single qubit ($D=d=2$) tomography are the Pauli operators and identity \cite{Nielsen_Chuang,Altepeter2005-QST,Usenko2024-QSTMaths},
    \begin{equation}
        \hat{\sigma}_X= \begin{pmatrix}
                    0&1\\
                    1&0
                  \end{pmatrix},
            \quad
        \hat{\sigma}_Y= \begin{pmatrix}
            0&-i\\
            i&0
            \end{pmatrix},
            \quad
            \hat{\sigma}_Z= \begin{pmatrix}
                1&0\\
                0&-1
                \end{pmatrix},
            \quad\mathrm{and}\quad
        \hat{\sigma}_I= \begin{pmatrix}
            1&0\\
            0&1
            \end{pmatrix},
        \label{eqn:pauli_mats}
    \end{equation}
    which can be linearly combined to form the density matrix, $\rho$, of any state $\ket{\psi}$
    \begin{equation}
        \rho = \ket{\psi}\bra{\psi} = \frac{1}{2}\sum_{i}S_i\hat{\sigma}_i.
        \label{eqn:single_qubit_rho}
    \end{equation}
    The parameters $S_i$, Stokes parameters, are calculated from a sum of the eigenvalues of $\hat{\sigma}_i$, $\lambda_{i,k}$, multiplied by their probability of occurring, $P_{i,k}$:
    \begin{equation}
        S_{i} = \sum_{k}\lambda_{i,k} P_{i,k}.
    \end{equation}
    In the case of the identity operator $\hat{\sigma}_I$, $\lambda_{I,0}=\lambda_{I,1}=1$, meaning it is always true that $S_I=1$ as this corresponds to all probabilities summing to 1. As a result, the density matrix construction is often shown with $\hat{\sigma}_I$ outside the sum.
    We can obtain the other $P_{i,k}$ experimentally by projecting the state $\rho$ onto the eigenstates, $\ket{e_{i,k}}$, of $\hat{\sigma}_i$, giving $P_{i,k}=Tr(\rho\ket{e_{i,k}}\bra{e_{i,k}})$. The operators $\hat{\Pi}_{i}$ that project on to the 2 eigenstates of each Pauli operator in Eq. \ref{eqn:pauli_mats} and their associated eigenvalues are
    \begin{equation}
        \begin{split}
        \hat{\Pi}_X= \frac{1}{\sqrt{2}}\begin{pmatrix}
                    1&1\\
                    1&-1
                  \end{pmatrix},\quad \lambda_{X,0}=1,\quad\lambda_{X,1}=-1;
            \\
        \hat{\Pi}_Y= \frac{1}{\sqrt{2}}\begin{pmatrix}
            1&i\\
            1&-i
            \end{pmatrix},\quad \lambda_{Y,0}=1,\quad\lambda_{Y,1}=-1;
            \\
            \hat{\Pi}_Z= \begin{pmatrix}
                1&0\\
                0&1
                \end{pmatrix},\quad \lambda_{Z,0}=1,\quad \lambda_{Z,1}=-1.
        \end{split}
        \label{eqn:qubit_projectors}
    \end{equation}
    We can note from the construction of $\hat{\Pi}_i$ that the top row of each matrix corresponds to the $\lambda_{i,0}$ eigenstate and the bottom row to the $\lambda_{i,1}$ eigenstate.
    To simplify the notation for the projectors onto the $\hat{\sigma}_{i}$ eigenstates, we will define $\ket{e_{i,k}}\bra{e_{i,k}}=\hat{\Pi}_{i,k}$ and the outcome of projecting $\rho$ onto the Pauli eigenstates is
    \begin{equation}
        \hat{\Pi}_i\rho = \begin{pmatrix}
            P_{i,0}\\P_{i,1}
        \end{pmatrix} = 
        \begin{pmatrix}
            Tr(\hat{\Pi}_{i,0}\psi)\\Tr(\hat{\Pi}_{i,1}\psi)
        \end{pmatrix}.
    \end{equation}
    Therefore, to complete Eq. \ref{eqn:single_qubit_rho} we perform 3 projective measurements each with 2 measured outcomes.
    It is important to distinguish between two ways of thinking about projective measurements: the $d$ projectors $\hat{\Pi}_m$ that each project onto the $n\in[0,d-1]$ states to give $d$ corresponding probabilities (this is the matrix we set our interferometer to perform); and the $d^2-1$ operators $\hat{\Gamma}_{p}=\ket{e_{p}}\bra{e_{p}}$ which project onto single states with probability $P_p=Tr(\hat{\Gamma}_p\rho)$, where
    \begin{equation}
        p=n+md
        \label{eqn:p_def}.
    \end{equation}
    In this notation, $m$ denotes the measurement basis and $n$ denotes the state within that basis. 
    
    Experimentally, for a photonic single-qubit state, we are setting our reconfigurable interferometer to perform the three projections of Eq. \ref{eqn:qubit_projectors} and measuring the number of single photon counts we get at the two outputs of the interferometer. Normalising the counts gives us $P_{i,0}$ and $P_{i,1}$. Performing these projections for each $\hat{\sigma}_i$ gives us all of the Stokes parameters, allowing us to construct the density matrix with Eq. \ref{eqn:single_qubit_rho}.

    \subsubsection{Multiple qubits}\label{Multiple qubits}
    We can extend Eq. \ref{eqn:single_qubit_rho} to $N$ qubits \cite{Nielsen_Chuang,Altepeter2005-QST}
    \begin{equation}
        \rho = \frac{1}{2^N}\sum_{i_1,i_2 \dots i_N}S_{i_1,i_2 \dots i_N}\hat{\sigma}_{i_1}\otimes\hat{\sigma}_{i_2}\otimes\dots\otimes\hat{\sigma}_{i_N}
    \end{equation}
    with the associated $N$-qubit Stokes parameters being
    \begin{equation}
        S_{i_1,i_2 \dots i_n} = \sum_{k_1,k_2 \dots k_n=0}^{n}\lambda_{i,k_1}\lambda_{i,k_2}\dots\lambda_{i,k_N}P_{i,k_1,i,k_2 \dots i,k_N}.
    \end{equation}
    In this construction of $\rho$, all projections for the $N$ qubits are applied locally. For measurements when all but one $\hat{\sigma}_{i_n}$ are identity, this corresponds to the local tomography of that particle.

    The probabilities in $N>1$ cases come from normalised coincidences between outputs of the different qubits (as opposed to the number of single photon counts like with the $N=1$ case). For example, for the measurement of two qubits $A$ and $B$, the probability $P_{X,0,Y,0}$ is obtained from applying the $\hat{\Pi}_X$ projector to $A$ and the $\hat{\Pi}_Y$ projector to $B$, and measuring coincidences between the zeroth mode of $A$ and the zeroth mode of $B$.
    
    \subsubsection{Single qudit}
    Up to this point, all measurements have been on $d=2$ systems, meaning the density matrix can be constructed using Pauli operators. As we increase the dimension we will need to consider what measurement basis to use. A set of operators that span the Hilbert space of a $d$-dimensional system can be obtained from the generalisation of the Pauli matrices, $\hat{\sigma}_i^{\{d\}}$ \cite{Bertlmann2008-GellMannGen}. We then use the same density matrix construction as in Eq. \ref{eqn:single_qubit_rho} using this set of operators. There are $d^2-1$ operators $\hat{\sigma}_i^{\{d\}}$, excluding identity, hence requiring $d^2-1$ projections, $\hat{\Pi}_i$, to achieve QST this way.
    QST is often performed using projectors of the generalised Pauli operators \cite{Wang2018-16D, Llewellyn2020-Thesis, Bao2023-VeryLargeScale}, however, this is not optimal in terms of the number of measurements required.
    The most measurement-efficient way of performing QST is using mutually unbiased bases (MUBs) \cite{Wootters1989-MUBTomo, Filippov2011-MUBTomo,Lima2011-MUBTomo}. A pair of bases $\hat{M}_l$, $\hat{M}_m$ are said to be mutually unbiased if, when a state is prepared from one basis and measured in another ($l\neq m$), all outcomes are equally likely \cite{Designolle2019-MUBs}. For integer power of prime number $d$ values, there are $d+1$ MUBs spanning the full Hilbert space \cite{Ding2017-MCFQKDChipToChip}. These constitute the projections $\hat{\Pi}_m$ experimentally performed, with each row of the MUB corresponding to a basis state and the normalised measurement being the probabilities of each basis state, as before.
    This set of $\hat{\Pi}_m$ projectors scales linearly with $d$, instead of quadratically compared to $d$-dimensional Pauli operator projections.
    It can be noted that the set of projectors for $\hat{\sigma}_{X}, \hat{\sigma}_{Y}, \hat{\sigma}_{Z}$ form a set of MUBs for $d=2$, so measuring a state using the Pauli operators is equivalent to performing MUB tomography for $d=2$.

    We will re-form and generalise Eq. \ref{eqn:single_qubit_rho} as \cite{Thew2002-QuditQST, Agnew2011-QuditQST}
    \begin{equation}
        \rho = \frac{1}{d}\hat{I}+\frac{1}{d}\sum_{i=1}^{d^2-1}S_i\hat{\sigma}_i^{\{d\}}.
        \label{eqn:single_qudit_rho}
    \end{equation}
    
    Now we have decided our set of measurements to be a set of MUBs, we need to consider how to express $\rho$ in terms of the set of projectors $\hat{\Gamma}_{p}$ and their probabilities $P_p$. To do this, we can define each $\hat{\sigma}_i^{\{d\}}$ as a linear combination of $\hat{\Gamma}_{p}$ where the coefficients $C_{p}$ are given by $Tr(\hat{\Gamma}_{p}\hat{\sigma}_i^{\{d\}})$,
    \begin{equation}
        \sigma_i^{\{d\}} = \sum_{p=1}^{d^2-1}C_{i,p}\hat{\Gamma}_{p}.
    \end{equation}
    This allows us to evaluate Eq. \ref{eqn:single_qudit_rho}, using the probabilities $P_p$ obtained form measuring the set of MUBs.
    The explicit reconstruction of a qudit measured with MUBs then is
    \begin{equation}
        \rho = \frac{1}{d}\hat{I}+\frac{1}{d}
        \sum_{i=1}^{d^2-1}
        \sum_{p=1}^{d^2-1}Tr(\hat{\Gamma}_{p}\hat{\sigma}_i^{\{d\}}) P_p\hat{\Gamma}_{p}.
        \label{eqn:full_single_qudit_rho}
    \end{equation}

    \subsubsection{Multiple qudits}
    Now that we know how to construct a density matrix using MUB measurements, we can extend this to construct the state of $N$ qudits. The measurement principle outlined here is based on performing separable measurements - i.e. projections are performed locally on each qudit.
    This is different from the approach of \cite{Wootters1989-MUBTomo} and \cite{Adamson2010-TwoQubitQSTNonSep} where non-separable MUB measurements are performed.
    The former theoretically outlines state reconstruction with MUB measurements while the latter presents experimental demonstration of multi-particle state reconstruction using MUB measurements.
    Such approach for $N$-particle measurement uses Eq. \ref{eqn:full_single_qudit_rho}, where the dimension of the MUB projectors is defined as the size of the joint state space 
    $d \rightarrow d^N$ and the $d^N+1$ MUBs are measured on the joint system which includes non-local measurements on the qudits. While this choice of measurement set constitutes the minimum number of measurements required to carry out tomography of $N$ qudits of dimension $d$ \cite{Wootters1989-MUBTomo}, it is not always possible to perform non-separable measurements, for example when the qudits are spatially disparate, as in this work. In the following, we outline how to linearly construct the density matrix of a state using combinations of d-dimensional MUBs measured locally on the $N$ qudits, totalling $(d+1)^N$ measurements, which constitutes the minimum number of separable measurements required to perform QST on $N$ qudits \cite{Usenko2024-QSTMaths}.

    When compared to the single-qudit case, our basis of projectors $\hat{\Pi}$ will go from $d^2-1$ to $d^{2N}-1$. To simplify notation, we will explicitly outline the two-qudit version here as this is relevant for the results presented in this work and many other two-photon experiments, but the principle can be extended to more qudits.
    The projection measurements performed on our qudits $a$ and $b$ are $\hat{\Pi}_a$ and $\hat{\Pi}_a$, meaning the joint measurement on the two-qudit state $\rho$ is $\hat{\Pi}_a\otimes\hat{\Pi}_b$. These joint measurements correspond to $d^{4}-1$ state projectors $\hat{\Gamma}_{a,b}$. We can break these down into
    \begin{equation}
        \begin{split}
        \hat{\Gamma}_{0,0} &= \hat{I}\otimes\hat{I}
        \quad \quad 1 \text{ measurement}
        \\
        \hat{\Gamma}_{a,0} &= \hat{\Gamma}_{a}\otimes\hat{I}
        \quad \text{ } \text{ } d^2-1 \text{ measurements}
        \\
        \hat{\Gamma}_{0,b} &= \hat{I}\otimes\hat{\Gamma}_{b}
        \quad \text{ }\text{ } d^2-1 \text{ measurements}
        \\
        \hat{\Gamma}_{a,b} &= \hat{\Gamma}_{a}\otimes\hat{\Gamma}_{b} \quad (d^2-1)(d^2-1) \text{ measurements}
        \end{split}
    \end{equation}
    where the top row is the identity, the middle two rows of projectors perform local tomography, and the bottom row captures the correlations between the two qudits.

    We can then extend the construction of Eq. \ref{eqn:full_single_qudit_rho} to two qudits \cite{Usenko2024-QSTMaths}
    \begin{equation}
        \rho = \frac{1}{d}\hat{\Gamma}_{0,0}+
        \frac{1}{d}\sum_{i_1,i_2}\sum_{a=1}^{d^2-1}\sum_{b=1}^{d^2-1}
        P_{a,b}(C_{i_1,i_2,a,0}\hat{\Gamma}_{a,0}+
        C_{i_1,i_2,a,0}\hat{\Gamma}_{0,b}+
        C_{i_1,i_2,a,b}\hat{\Gamma}_{a,b}),
        \label{eqn:full_two_qudit_rho}
    \end{equation}
    where $C_{i_1,i_2,a,b}=Tr(\hat{\Gamma}_{a,b}\hat{\sigma}_{i_1}^{\{d\}}\otimes\hat{\sigma}_{i_1}^{\{d\}})$. The first term in the sum is the local tomography of qudit $a$, second the local tomography of qudit $b$ and final term captures the correlations between the two. This can be extended to any number of qudits by: replacing $P_{a,b}$ with probabilities from coincidences between the $N$ qudits; including $N$ terms in the sum for the local tomography of each qudit; including all combinations of tensor products of $\hat{\Gamma}$ and $\hat{\sigma}$ projectors and operators.
    An excellent mathematical description of linear quantum state reconstruction of systems of any dimension and number of particles, using any complete measurement basis can be found in \cite{Usenko2024-QSTMaths}, including the linear construction of high-dimensional, multi-qudit states using separable MUB measurements.

    \subsection{Performing measurements}\label{sec:performing_measurements}
    Knowing how to linearly construct a density matrix informs the choice of measurements we make. As discussed, in this work we perform measurements in mutually unbiased bases to minimise the number of measurements required. In order to experimentally perform measurements on a pair of $a$ and $b$, we configure on-chip unitaries $\hat{A}$ and $\hat{B}$ to apply the projection operators $\hat{\Pi}_a$ and $\hat{\Pi}_b$ to the idler and signal photons respectively. The unitaries are implemented on-chip with beamsplitters and phase shifters, where the tunability of the phase shifters allows us to reconfigure the circuit to realise difference matrices.
    We work out the phases to set in the circuit to implement different projectors using a minimisation algorithm that minimises the difference between the desired projection matrix and the on-chip unitary matrix, parametrised by the phase shifters.
    We then set these phases on-chip and measured coincidences between the outputs of Alice and Bob. Normalised coincidence counts between a pair of outputs correspond to the probability of the associated eigenstate of $\hat{\Pi}_a \otimes \hat{\Pi}_b$.

    \subsection{Physical state estimation}
    The above linear reconstruction allows for the construction of a density matrix from the measured projections and probabilities. However, this reconstruction is unlikely to represent a \textit{physical} density matrix according to the properties of a density matrix outlined at the start of Section \ref{sec:QST} due to noise in experimental measurements. Therefore, we must perform state estimation, which finds the physical density matrix that most closely produces the experimentally obtained measurement outcomes.
    For this, we need a parameterised matrix which enforces the rules of a physical density matrix. Any $D\times D$ matrix that can be written in the form $\hat{T}^\dagger\hat{T}$ is Hermitian \cite{Altepeter2005-QST} and to enforce normalisation we can divide by $Tr(\hat{T}^\dagger\hat{T})$, therefore any matrix that can be written as 
    \begin{equation}
        \rho_{phy} = \frac{\hat{T}^\dagger\hat{T}}{Tr(\hat{T}^\dagger\hat{T})}
    \end{equation}
    represents a physical density matrix \cite{Altepeter2005-QST}, $\rho_{phy}$. As we wish to invert $\hat{T}$, it is convenient to choose its form to be a (lower or upper) triangular matrix,
    \begin{equation}
        \hat{T}=
        \begin{pmatrix}
            t_{1,1}              & 0                    & \dots&0\\
            t_{2,1}+it_{2+D,1+D} & t_{2,2}              & \dots  &0\\
            \vdots               & \vdots               & \ddots &\vdots \\
            t_{D,1}+it_{2D,1+D} & t_{D,2}+it_{2D,2+D}   & \dots  &t_{D,D}\\
        \end{pmatrix}
    \end{equation}
    parameterised by $t$ parameters, of which there are $d^{2N}$. In order to find the $t$-parameters we can minimise the difference between the statistics from the linearly constructed density matrix, $\rho_{lin}$, and $\rho_{phy}$
    \begin{equation}
        \mathcal{L}(t_{1,1},t_{1,2}...t_{D,D}) = \sum_{p} (Tr(\hat{\Gamma}_{p}\rho_{phy})-Tr(\hat{\Gamma}_{p}\rho_{lin}))^2.
        \label{eqn:cost_function}
    \end{equation}

    \subsection{State metrics}
    From the constructed density matrix we can now quantify fidelity, entanglement, and dimensionality. Fidelity, $\mathcal{F}$, is a metric for the closeness of two density matrices. The fidelity of the experimentally measured state $\tilde{\rho}$ when compared to the ideal state $\rho$ is defined as \cite{Altepeter2005-QST}
    \begin{equation}
        \mathcal{F} = \left(Tr\left(\sqrt{\sqrt{\rho}\tilde{\rho}\sqrt{\rho}}\right)\right)^2.
    \end{equation}
    
    Entanglement entropy, $\mathcal{E}$, can be used to quantify the degree of entanglement between two particles $A$ and $B$. $\mathcal{E}$ is calculated using the reduced density matrix of $A$, $\rho_A$, or $B$, $\rho_B$. In particular their von Neumann entropy is defined as \cite{Altepeter2005-QST}
    \begin{equation}
        \mathcal{E} = -Tr(\rho_A \log_d(\rho_A)) = -Tr(\rho_B \log_d(\rho_B)).
    \end{equation}
    $\mathcal{E}=0$ indicates a state with no entanglement, corresponding to a product state. In contrast a non-zero value of $\mathcal{E}$ indicates entanglement in present in the state $\rho$, where $\mathcal{E}=1$ corresponds to a maximally entangled state.

    We can certify the dimension of the qudits by quantifying the minimum $d$ required to describe the measured state using a dimension witness $\lceil{\mathcal{D}}\rceil$ defined as \cite{Wang2018-16D}
    \begin{equation}
        \lceil{\mathcal{D}}\rceil = \frac{1}{\left\lfloor \sum_{n_b,n_{b'}}\left(\sum_{n_a}\sqrt{P_{a,b}P_{a,b'}}\right)^2 \right\rfloor}
    \end{equation}
    where $n_x$ denotes the state within the basis $m_x$, and $x=n_x + dm_x$ as in Eq. \ref{eqn:p_def} (where $x\in [a,b,b']$). This allows us to calculate $\lceil{\mathcal{D}}\rceil$ by measuring the first qubit in one basis ($\hat{\Pi}_{m_a}$) and the second in two different bases ($\hat{\Pi}_{m_b}$ and $\hat{\Pi}_{m_{b'}}$, where $b\neq b'$).

    \subsection{Error calculation}
    Errors on the state metrics are calculated from Monte-Carlo simulations using the measured statistics. This is done by re-sampling the measured probabilities from a distribution defined by the error on the measured probabilities, in this case taken as a Poissonian distribution where the mean is the measured probability and the standard deviation is given by the error on the measured probability \cite{Adcock2019-thesis}. Error on the measured probability is calculated from the error in measured counts which is defined as a convolution of poissonian counting statistics for the raw coincidence counts and standard error on the mean of the measured detector channel efficiencies (outlined in Section \ref{sec:dce}). The re-sampled statistics are used to construct a new physical density matrix for which we calculate the state metrics. This re-sampling and state reconstruction process is repeated 1,000 times sufficient for standard deviation for each metric to converge.

    \subsection{Tomography workflow summary}
    Section \ref{sec:QST} so far provides a practical guide to carrying out QST, summarised as follows:
    \begin{enumerate}
        \item Selecting the measurements to be performed.
        \item Calculating the phases to set in the on-chip unitaries to perform the measurements.
        \item Experimentally preparing the state and performing the projections by applying unitaries to obtain probabilities.
        \item Linearly constructing a non-physical density matrix using the measured probabilities.
        \item Performing state estimation to find the physical state that most closely resembles the linearly constructed state.
        \item Calculating state metrics, in this case: fidelity, entanglement entropy, and dimension witness.
        \item Performing Monte-Carlo simulations and repeating state construction and estimation to gain distributions of the state metrics to calculate errors.
    \end{enumerate}

\subsection{Generalised Pauli projectors only have up to 2-mode superpositions}
Multi-mode phase stabilisation enables us to perform quantum state tomography with mutually-unbiased basis measurements which require fewer measurements for completeness than using generalised Pauli operator projectors. Typical qudit tomography demonstrations use Pauli operators as the eigenstates contain no more than two modes in superposition, and therefore require phase coherence of just two modes - as we show in the following.

The three groups of generalised Pauli operators for dimension $d$ are constructed as follows \cite{Bertlmann2008-GellMannGen}:
\begin{equation} \label{ggms}
    \text{Symmetric: }
        \hat{A}_{jk} \;=\; \ketbra{j}{k} + \ketbra{k}{j}, \quad 0 \leq j < k \leq d-1 \;,
    \end{equation}
\begin{equation} \label{ggma}
    \text{Anti-symmetric: }
        \hat{A}_{jk} \;=\; -i \ketbra{j}{k} + i \ketbra{k}{j}, \quad 0 \leq j < k \leq d-1 \;,
    \end{equation}
\begin{equation} \label{ggmd}
        \text{Diagonal: }
        \hat{A}_l \;=\; \sqrt{\frac{2}{(l+1)(l+2)}} \left( \sum_{j=0}^{l}
        \ketbra{j}{j} - (l+1)\ketbra{l+1}{l+1}\right), \quad 1 \leq l \leq d-2 \;.
    \end{equation}

In order to prove that generalised Pauli operator projectors depend on no greater than 2 modes in superposition, we will show that the eigenstates, $\ket{\psi_\lambda}=\sum_{n=0}^{d-1}\alpha_n\ket{n}$, of the generalised Pauli operators consist of no more than two-basis states using the eigenstate condition $\hat{A}\ket{\psi_\lambda}=\lambda\ket{\psi_\lambda}$, where $\lambda$ are the eigenvalues.

In the \textit{symmetric} case:
\begin{equation} \label{symmetric}
\begin{split}
\hat{A}_{jk}\ket{\psi_\lambda} & = \ketbra{j}{k}\sum_{n=0}^{d-1}\alpha_n\ket{n}+\ketbra{j}{k}\sum_{n=0}^{d-1}\alpha_n\ket{n}\\
 & = \alpha_k\ket{j} + \alpha_j\ket{k}
\end{split}
\end{equation}
\begin{equation}
    \therefore \quad \alpha_k\ket{j}+\alpha_j\ket{k} = 
    \lambda\ket{\psi_\lambda} = \lambda\sum_{n=0}^{d-1}\alpha_n\ket{n}=
    \lambda\alpha_j\ket{j}+\lambda\alpha_k\ket{k}.
\end{equation}
As such, we can calculate the eigenvalues
\begin{equation}
\begin{split}
    \alpha_k=\lambda\alpha_j,\; \alpha_j=\lambda\alpha_k 
    &\implies \lambda=\pm1 \,, \\
    \lambda\alpha_n=0 &\implies \lambda=0 \; [n\neq j,k] \,,
    \end{split}
\end{equation}
and associated eigenvectors
\begin{equation}
\begin{alignedat}{2}
    \lambda=1 &\implies \alpha_k=\alpha_j   & &\implies \ket{\psi_1}=\alpha_j(\ket{j}+\ket{k})  \,,\\
    \lambda=-1 &\implies \alpha_k=-\alpha_j & &\implies \ket{\psi_{-1}}=\alpha_j(\ket{j}-\ket{k})  \,,\\
    \lambda=0 &\implies \alpha_n=\alpha_n, \, \alpha_j,\alpha_k=0   & &\implies \ket{\psi_0}=\alpha_n\ket{n} \; [n\neq j,k] \,,\\
    \end{alignedat}
\end{equation}
where no eigenvector has more than two modes in superposition. The eigenvalue $\lambda=0$ is $(d-2)$-fold degenerate, meaning the associated $d-2$ eigenvectors can be written in any orthonormal basis, which we chose for simplicity to be the computational basis, $\{\ket{n} \text{ for all } n\neq j,k\}$. This is true for the symmetric, anti-symmetric, and diagonal cases.

In the \textit{anti-symmetric} case:
\begin{equation} \label{antisymmetric}
\begin{split}
\hat{A}_{jk}\ket{\psi_\lambda} & = -i\ketbra{j}{k}\sum_{n=0}^{d-1}\alpha_n\ket{n} + i\ketbra{j}{k}\sum_{n=0}^{d-1}\alpha_n\ket{n}\\
 & = -i\alpha_k\ket{j} + i\alpha_j\ket{k}
\end{split}
\end{equation}
\begin{equation}
    \therefore \quad -i\alpha_k\ket{j} + i\alpha_j\ket{k} = \lambda\sum_{n=0}^{d-1}\alpha_n\ket{n}=
    \lambda\alpha_j\ket{j}+\lambda\alpha_k\ket{k}.
\end{equation}
As such, we can calculate the eigenvalues
\begin{equation}
\begin{split}
    -i\alpha_k=\lambda\alpha_j,\; i\alpha_j=\lambda\alpha_k 
    &\implies \lambda=\pm1 \,, \\
    \lambda\alpha_n=0 &\implies \lambda=0 \; [n\neq j,k] \,,
    \end{split}
\end{equation}
and associated eigenvectors
\begin{equation}
\begin{alignedat}{2}
    \lambda=1 &\implies \alpha_k=i\alpha_j   & &\implies \ket{\psi_1}=\alpha_j(\ket{j}+i\ket{k})  \,,\\
    \lambda=-1 &\implies \alpha_k=-i\alpha_j & &\implies \ket{\psi_{-1}}=\alpha_j(\ket{j}-i\ket{k})  \,,\\
    \lambda=0 &\implies \alpha_n=\alpha_n, \, \alpha_j,\alpha_k=0   & &\implies \ket{\psi_0}=\alpha_n\ket{n} \; [n\neq j,k] \,,\\
    \end{alignedat}
\end{equation}
where no eigenvector has more than two modes in superposition.

And, finally, in the \textit{diagonal} case:
\begin{equation} \label{diagonal}
\begin{split}
\hat{A}_{l}\ket{\psi_\lambda} & = \sum_{j=0}^{l}\ketbra{j}{j}\sum_{n=0}^{d-1}\alpha_n\ket{n} - (l+1)\ketbra{l+1}{l+1}\sum_{n=0}^{d-1}\alpha_n\ket{n}\\
 & = \sum_{j=0}^{l}\alpha_j\ket{j} - (l+1)\ket{l+1}
\end{split}
\end{equation}
\begin{equation}
    \therefore \quad \sum_{j=0}^{l}\alpha_j\ket{j} - (l+1)\ket{l+1} = \lambda\sum_{n=0}^{d-1}\alpha_n\ket{n}=
    \lambda\sum_{j=0}^{l}\alpha_j\ket{j}+\lambda\alpha_{l+1}\ket{l+1}.
\end{equation}
As such, we can calculate the eigenvalues
\begin{equation}
\begin{split}
    \lambda\alpha_n=\alpha_j,\;                    &\implies \lambda=1 \,, \\
    \lambda\alpha_{l+1}=-(l+1)\alpha_{l+1}  &\implies \lambda=-(l+1) \,, \\
    \lambda\alpha_n=0                       &\implies \lambda=0 \; [n > l+1] \,,
    \end{split}
\end{equation}
and associated eigenvectors
\begin{equation}
\begin{alignedat}{2}
    \lambda=1 &\implies \alpha_n=\alpha_j, \, \alpha_{l+1}=0   & &\implies \ket{\psi_1}=\alpha_j\ket{j}  \,,\\
    \lambda=-(l+1) &\implies -(l+1)\alpha_n=\alpha_n, \, \alpha_{l+1}=1 & & \implies \ket{\psi_{-(l+1)}}=-(l+1)\ket{l+1}  \,,\\
    \lambda=0 &\implies \alpha_j=1, \, \alpha_{l+1}=0, \, \alpha_n=\alpha_n  & &\implies \ket{\psi_0}=\alpha_n\ket{n} \; [n\neq l+1] \,,\\
    \end{alignedat}
\end{equation}
where, again, no eigenvector has any modes in superposition.

Intuitively this makes sense as all generalised Pauli matrices are permutations of the $d=2$ Pauli operators embedded in $d$-dimensional matrices. We can think of their projectors as probing many, smaller parts of the quantum state, as opposed to mutually-unbiased basis measurements which probe larger parts of the state per measurement, hence requiring fewer measurements.

\bibliographystyle{unsrt}
\bibliography{HDEntDis_bib}

\end{document}